Review

# Interdiffusion in group IV semiconductor material systems: applications, research methods and discoveries

**Guangrui (Maggie) Xia**

Department of Materials Engineering, University of British Columbia, Vancouver, BC V6T 1Z4, Canada

guangrui.xia@ubc.ca

## Abstract
Group IV semiconductor alloys and heterostructures such as SiGe, GeSn, Ge/Si and SiGe:C have been widely used and under extensive research for applications in major microelectronic and photonic devices. In the growth and processing of these materials, nanometer scale interdiffusion happens that are generally undesirable for device performance. With higher Ge molar fractions and higher compressive strains, Si-Ge interdiffusion can be much faster than dopant diffusion. However, Si-Ge interdiffusion behaviors have not been well understood until recent years. Much less studies are available for GeSn. This review starts with basic properties and the applications of major group IV semiconductors, and then reviews the progress made so far on Si-Ge and Ge-Sn interdiffusion behaviors. Theories, experimental methods, design and practical considerations are discussed together with the key findings in this field.

**Keywords:**
Group IV semiconductors, interdiffusion, modelling, SiGe materials and devices

# 1 Introduction and background

In the periodic table of elements, non-synthetic group IV elements include carbon (C), silicon (Si), germanium (Ge), tin (Sn) and lead (Pb). Among these elements, Si and Ge are the two most important elemental semiconductors. Specifically, Si has played a dominant and unreplaceable role in the semiconductor industry, which has revolutionized the world by enabling the Information Era. According to the product breakdown statistic data by World Semiconductor Trade Statistics (WSTS) 2017 report, Si-based semiconductor products including integrated circuits (ICs) and discrete components have occupied about 90% of the worldwide semiconductor market. Today, the infrastructure, processing, analytical techniques and equipment used in the semiconductor industry are highly centered for Si-based products.

Besides the elemental semiconductors, group IV elements form a wide range of important semiconductors, and have wide applications in metal oxide semiconductor field effect transistors (MOSFETs), heterojunction bipolar transistors (HBTs) and photonic devices (Fig. 1). They are in the forms of alloys (SiGe, SiGeC, GeSn, SiGeSn, GeC) and compounds (SiC). The incorporation of C in Si, SiGe and Ge can surpass the equilibrium solid solubility limits in temperature ranges without the formation of a second phase. For technology-relevant cases, C concentrations are commonly below 2 atomic percent (at.%). The term "$Si_{1-x}C_x$ alloy" or "Si:C solid-state solution" are both used. The latter one is used in this paper to avoid possible confusions with silicon carbide, which is a wide-bandgap semiconductor suitable for power applications. Carbon allotropes such as graphene and carbon nanotubes are also interesting materials for electronic applications, but will not be discussed here. This paper focuses on group IV semiconductors important for microelectronic and photonic applications, such as Si, SiGe, SiGe:C, Ge and GeSn.

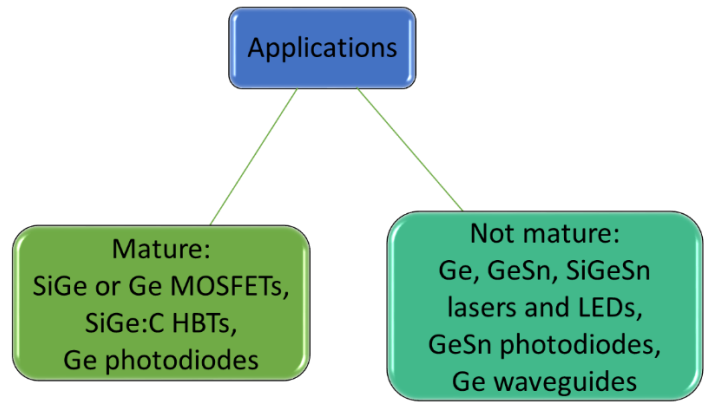

Fig. 1. (Color online) Major device applications of SiGe, Ge, GeSn, SiGeSn and SiGe:C.

## 1.1. The concepts, significance and necessity of interdiffusion research of group IV systems

Despite close material properties, SiGe, SiGe:C, Ge and GeSn are still different material systems than Si. For example, the lattice mismatch between Ge and Si causes direct growth of high quality epitaxial $Si_{1-x}Ge_x$ on Si difficult. Strain relaxation in lattice mismatched $Si_{1-x}Ge_x/Si_{1-y}Ge_y$ and $Ge_{1-x}Sn_x/Ge_{1-y}Sn_y$ epitaxial structures happens by misfit dislocations when the strained film thickness is over a critical thickness. The thermal expansion coefficient of Ge is about twice of that of Si, which introduces thermal stress upon cooling and heating. The epitaxial growth, defect formation, lattice-mismatch stress, thermal-mismatch stress and stress relaxation are all important considerations in the device processing, which have been studied intensively in the past few decades.

In high-end semiconductor devices, such as MOSFETs, bipolar transistors, lasers, and high efficiency solar cells, single crystalline semiconductors are commonly doped with at least one dopant element. During the growth and processing of the semiconductor materials, high temperature steps, such as oxidation, source/drain anneal and deposition, are unavoidable. From a mass transport point of view, when a host matrix changes from Si to SiGe or to SiGe:C, or even with a change in stress/strain, dopant diffusion changes. On top of that, dopants segregate at $A_{1-x}B_x/A_{1-y}B_y$ interfaces or with a gradual change in the composition. Interdiffusion changes the chemical concentration distributions directly, which impact the bandgap alignment, carrier transport, dopant diffusion and segregation behaviors. For example, Si-Ge interdiffusion is generally undesirable as sharp concentration differences are desired for quantum confinement or strain introduction. These three phenomena are significant topics for device structure design and materials processing, as dopant and alloy element distributions are critical factors in determining device characteristics and performance.

### 1.1.1 Concepts and examples

Interdiffusion or diffusion is a physical phenomenon in a mixture, where microscopic particles (atoms, ions, electrons, holes, molecules etc.) of the constituents have net movements due to their chemical potential gradients. The term "diffusion" and "interdiffusion" are both used. In some dilute solution cases, we can consider or approximate that one constituent moves in a fixed diffusion medium. In these cases, "diffusion" is commonly used. For example, common dopants in semiconductors are below 1 at.%, and the diffusion of dopants is called dopant diffusion instead of dopant interdiffusion. In other cases, especially when A and B have comparable concentrations, "interdiffusion" is a more precise term, as it stresses the fact that all constituents can have net movements.

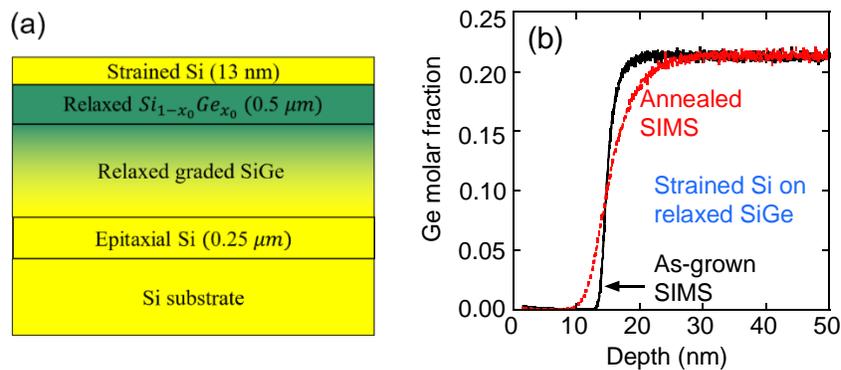

Fig. 2. An example of Si-Ge interdiffusion at a Si/Si$_{0.79}$Ge$_{0.21}$ interface. (a) The schematic structure with the buffer layer and the substrate also shown, and (b) the Ge molar fraction profile before and after annealing. The annealing condition was 920 °C for 60 min [1]. Figure reprinted from Ref. [1] with permission from AIP publishing.

Figure 2 gives an example of Si-Ge interdiffusion at a strained Si/relaxed $Si_{0.79}Ge_{0.21}$ interface for n-MOSFET applications [1]. The interface had a sharp Ge concentration step before the diffusion at 920 °C for 60 min, after which, the interface changed to an alloy region of about 10 nm thick (Fig. 2b). The interdiffusion introduces alloy scattering in the Si layer, which degrades electron mobility.

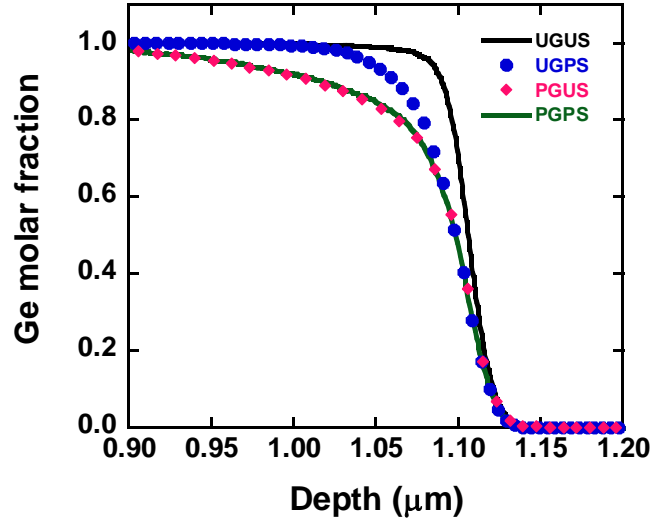

Fig. 3. Ge molar fraction profiles of Ge/Si interfaces after a defect annealing step, which was a thermal cycling step with an equivalent thermal budget of 725 °C for 64 min. Sample UGUS, UGPS, PGUS and PGPS stand for undoped Ge/undoped Si, undoped Ge/P-doped Si, P-doped Ge/undoped Si, and P-doped Ge/P-doped Si respectively [2]. Figure reprinted from Ref. [2] with permission from AIP publishing.

For Ge/Si structures, after a defect annealing step, which is a standard step right after the growth of Ge on Si by chemical vapour deposition (CVD), the interfaces are not ideal sharp interfaces (Fig. 3). Instead, due to the interdiffusion during the growth and defect annealing, the interfaces become SiGe alloy regions. For Ge-on-Si laser applications, n-type dopants, such as phosphorus (P), enhance the interdiffusion. The interdiffusion regions for undoped Ge/Si (UGUS) and P-doped Ge/Si (PGPS) are about 60 and 225 nm thick if we define the interdiffusion region is where $0.02 < x_{Ge} < 0.98$. If we define the interdiffusion region is where $0.05 < x_{Ge} < 0.95$, the alloy regions are about 45 and 159 nm thick for UGUS and PGPS respectively, which are significant portions of the Ge films, commonly a few hundred of nanometers, for Ge-on-Si lasers [2].

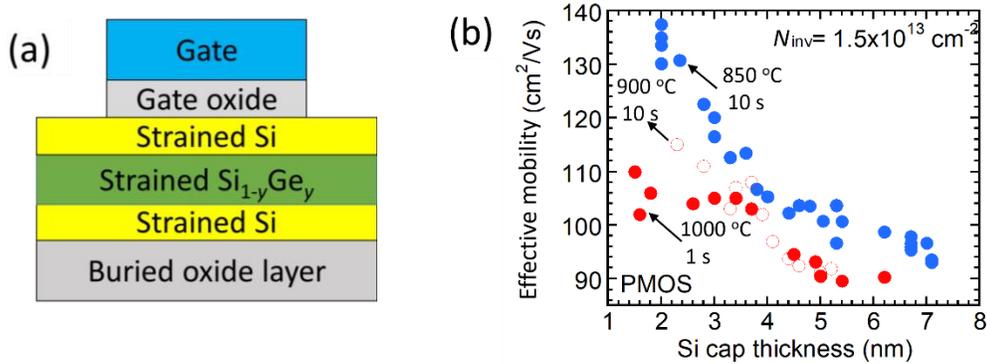

Fig. 4. An example of Si-Ge interdiffusion and its impact on the effective hole mobility of a p-type Si/Si$_{0.54}$Ge$_{0.46}$/Si heterostructure on insulation (HOI) p-type MOSFET. (a) The heterostructure has $y=0.46$ and the two strained Si layers have +1% tensile strain. (b) The impact of thermal annealing on the effective hole mobility. The Ge profiles measured by secondary ion mass spectrometry (SIMS) before and after rapid thermal annealing (RTA) showing interdiffusion [3]. Figure reprinted from Ref. [3] with permission from ECS.

1.1.2 Interdiffusion impacts on device performance

Four examples on MOSFETs, HBT and Ge lasers are discussed below to illustrate the significance of interdiffusion on semiconductor device performance. Figure 4 has an example of Si-Ge interdiffusion and its impact on the effective hole mobility of a p-type Si/Si$_{0.54}$Ge$_{0.46}$/Si heterostructure on insulation (HOI) MOSFET. In this design, the buried SiGe layer was the hole channel. Higher Ge concentration and compressive stress both help to enhance the hole mobility. After the source/drain activation annealing of 850 and 950 °C for 10 s, due to interdiffusion, the peak Ge concentration dropped from 50% to 46% and 38% respectively and the Ge peak widened. Therefore, the interdiffusion degraded the hole mobility at a high hole inversion charge density $N_{inv}$, especially for the devices with thinner Si cap thickness (Fig. 4b). The second example is on strained-Si/relaxed-SiGe n-type MOSFETs, implantation-enhanced Si-Ge interdiffusion degrades the electron mobility (Fig. 5) even for a low implant dose of $2.7\times10^{13}$ cm$^{-2}$. This implant condition is relevant for common sub-50 nm MOSFETs with halo and extension implants.

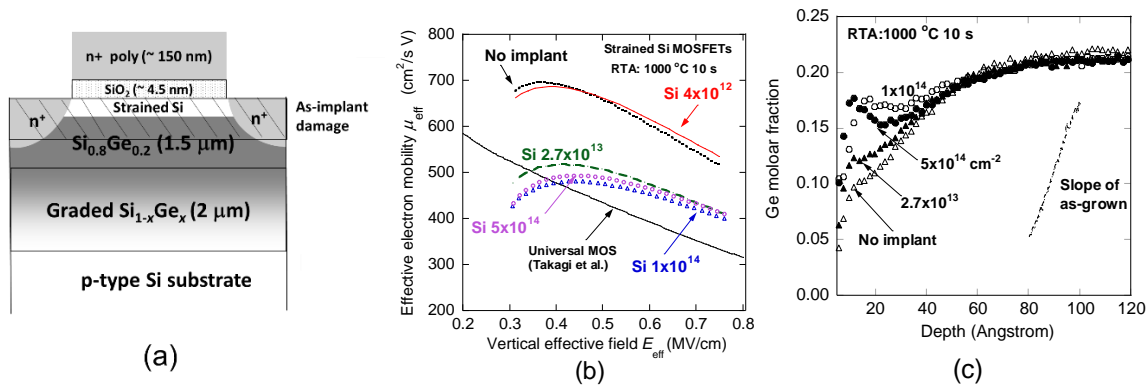

Fig. 5. An example of Si-Ge interdiffusion and its impact on the effective electron mobility of an n-type strained-Si/relaxed-Si$_{0.8}$Ge$_{0.2}$ MOSFET. (a) The structure of the MOSFET; (b) the dependence of the electron mobility on the ion implant dose; and (c) the SIMS Ge profiles illustrating the implant-enhanced interdiffusion, which brings Ge atoms

towards the original strained Si channel and degrades the mobility [4]. Figure reprinted from Ref. [4] with permission from IEEE.

The third example is on n-Ge/Si lasers. N-type doping with phosphorus (P) enhances the interdiffusion by a factor of 2 to 8 times depending on the germanium molar fraction [5, 6]. The photoluminescence of n-Ge was greatly reduced due to the interdiffusion during the defect annealing step (Fig. 6) [7]. After the defect annealing, there is a thick SiGe transition region between Ge and Si. The peak for direct transitions blue-shifted after annealing, and the PL showed an increase of the direct bandgap due to the SiGe alloying as seen in Fig. 6c. There is also a reduction in the fraction of direct transitions.

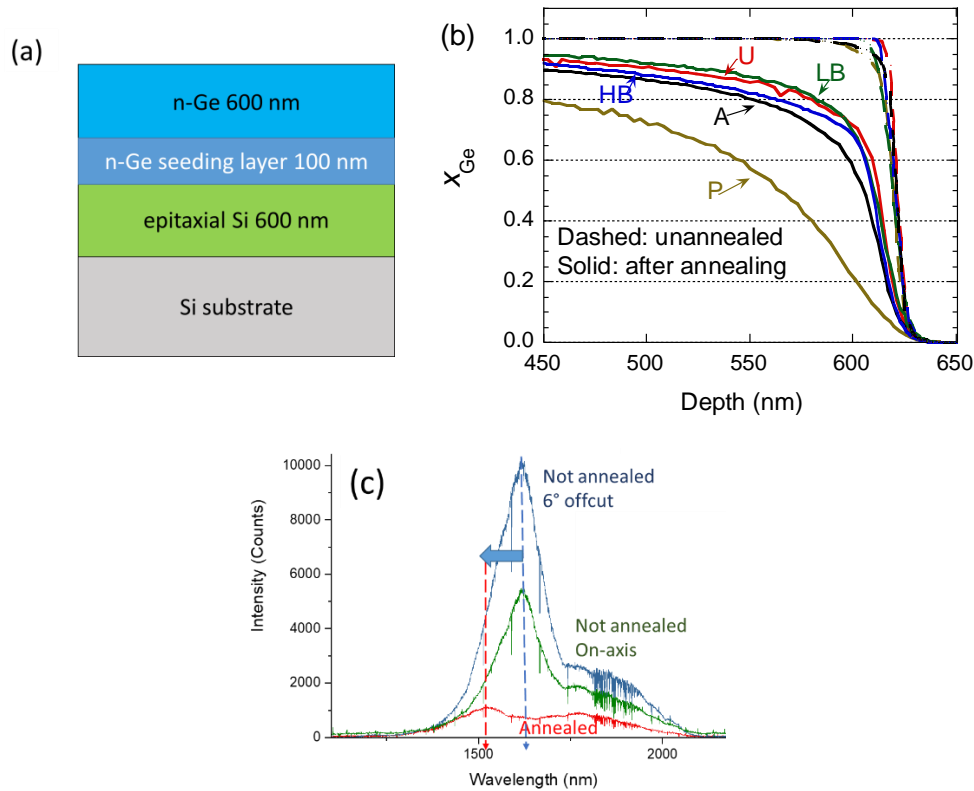

Fig. 6. (Color online) Si-Ge interdiffusion and its impact on the photoluminescence (PL) of P-doped Ge on Si. (a) Schematic structure of n-Ge/Si; (b) Ge profiles of 600 nm Ge on Si before and after defect annealing. U: undoped, LB: lower boron doped, HB: higher boron doped, P: P-doped and A: As-doped; (c) PL measurements of annealed P-doped Ge/Si with and without defect annealing. (a) and (b) reprinted from Ref. [6] with permission from Opitcal Society of America (OSA).

The fourth example is on SiGe HBTs. Higher Ge fractions ($x_{Ge}$) narrow the energy bandgap. The bandgap engineering with SiGe improves emitter injection efficiencies and provides an additional acceleration for electrons to transit through the base regions. These effects result in a significant increase in the unit current gain cut-off frequency $f_T$, which is the one of the most important figures-of-merit for HBT performance.

In the late 80's, IBM developed the SiGe HBT technology. In Ref. [8], 65 nm thick graded SiGe base with $x_{Ge}$ up to 11% was used instead of Si, which showed a 10X collector current enhancement and a 30% reduction in base transit time to 2.6 ps. These SiGe HBTs have peak $f_T$ of 45 GHz in comparison with 38 GHz of the Si counterparts. As shown in Fig. 7, Patton et al. [10] used graded Ge up to $x_{Ge}$ = 8% in the base regions. The total thickness of the graded Ge and the Ge peak was about 70 nm. SiGe HBTs achieved a $f_T$ of 75 GHz in comparison with 52 GHz from the Si counterparts.

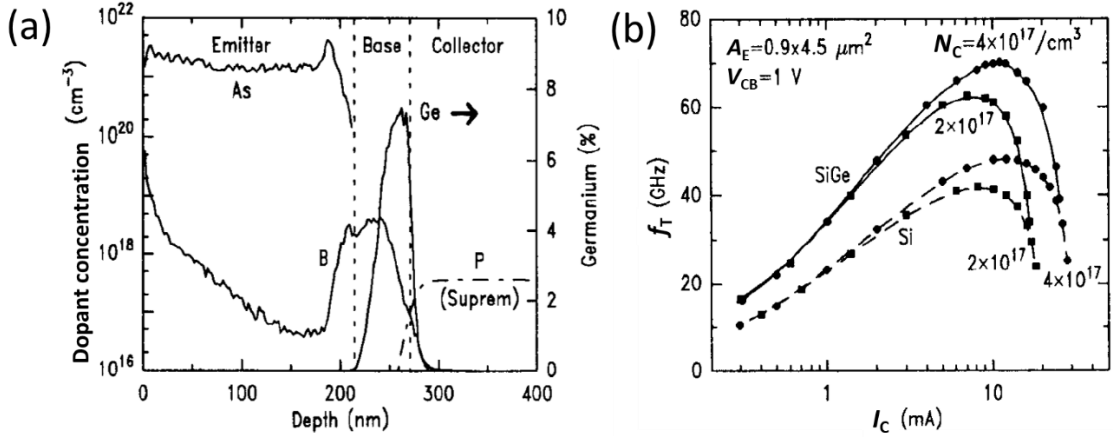

Fig. 7. Element profiles and performance of the SiGe HBTs developed in the late 80's: (a) SIMS profiles of boron, Ge and arsenic (As) concentrations of the emitter, base and collector. The Ge molar fraction is close to 8%, and the width of the SiGe layer is about 70 nm; (b) The $f_T$ of a SiGe HBT of the same work in comparison with its Si counterpart, showing a significant increase in $f_T$ over the full $I_C$ range [8]. Figure reprinted from Ref. [9] with permission from IEEE.

For SiGe HBTs, the SiGe base layers are generally strained. The bandgap of a strained SiGe film (at 300 K) depends on the Ge fraction $x$ as in the following equation [10]:
$$E_g = 1.17 - 0.94x + 0.34x^2 \text{ (eV)}. \quad (1)$$
The collector current $I_C$ depends on the bandgap change and Ge grading exponentially [10]:
$$I_{C,SiGe} = I_{C,Si} * \tilde{\gamma} * \tilde{\eta} * \frac{\Delta E_{g,Ge(grade)}}{kT} * \frac{\exp\left(\frac{\Delta E_{g,Ge(0)}}{kT}\right)}{1-\exp\left(\frac{-\Delta E_{g,Ge(grade)}}{kT}\right)}. \quad (2)$$

In the above equation, the term delta $\Delta E_{g,Ge(0)}$ is the Ge-induced bandgap narrowing at the emitter end of the base. $\Delta E_{g,Ge(grade)}$ is the grading of the Ge across the base. $\tilde{\gamma}$ and $\tilde{\eta}$ are the effective density of states and the minor electron diffusivity ratio, respectively. The addition of Ge in Si improves the emitter injection efficiency and provides an additional acceleration for electrons to transit through the base region, which increases $I_C$. This increase is one of the most effective ways to increase $f_T$. Therefore, even 1% Ge molar fraction change is important for the HBT design. In modern SiGe HBTs, commonly used $x_{Ge}$ of Ge peaks is about 20 at.%, and higher $x_{Ge}$ about 30 at.% has been actively in research.

As group IV alloys are widely used, when considering the interdiffusion process, we also need to consider the scale of the interdiffusion length, which is the characteristic length of interdiffusion.

Commonly, interdiffusion regions are within tens of nanometers in size. It is an important topic for SiGe, SiGe:C, GeSn layers or regions thinner or smaller than 100 nm. Such applications include modern FETs, HBTs and MQW-based devices. For group IV devices of micron scale and without n-type doping, such as Ge photodiodes and modulators, interdiffusion can be ignored. For n-doped Ge or SiGe with high compressive strains, due to the doping and strain impacts, interdiffusion can be greatly accelerated and become a problem, such as the cases for n-Ge lasers and SiGe clusters under high compressive strains for light emitting applications [11].

Due to the significance of the Ge distribution, Si-Ge interdiffusion models and parameters have been included in the most widely used semiconductor process simulation tools, such as Taurus Process$^{TM}$, TSUPREM-4$^{TM}$ and Sentaurus Process$^{TM}$ by Synopsys, since 2004. Interdiffusion was modelled as dopant diffusion in Taurus Process$^{TM}$ in 2004 and later versions and in Sentaurus Process$^{TM}$'s 2007 and later versions. In 2014, more sophisticated interdiffusion models were adopted by Sentaurus Process$^{TM}$.

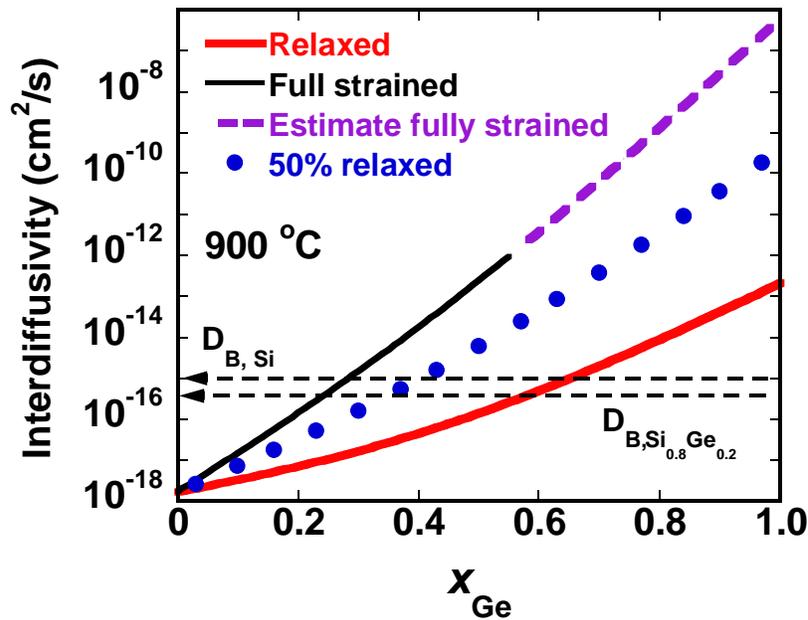

Fig. 8. (Color online) Si-Ge interdiffusivity at 900 °C for the whole Ge fraction range in comparison with boron (B) diffusivity in Si and Si$_{0.8}$Ge$_{0.2}$. The term "strained" and "relaxed" refer to the SiGe strain status on a relaxed Si substrate. Due to the very thin critical thickness (a few nanometer and below), fully strained high-Ge alloys coherent to a relaxed Si are not common, and are denoted with a dashed line. The interdiffusivity models used are to be discussed in Session 2.5 and 3.3.

It is interesting to compare dopant diffusivities and interdiffusivities. Dopants are impurities in semiconductors, which can contribute free electrons or free holes to the host semiconductors. Common doping ranges are from $1\times10^{15}$ to $2\times10^{20}$ cm$^{-3}$ for SiGe, SiGe:C and GeSn alloys. Doping means the addition of dopants for the purpose of carrier introduction or engineering.

Previously, Si-Ge interdiffusion was not as important as dopant diffusion, due to the much slower interdiffusivity in SiGe with lower Ge contents and lower compressive strain levels. In Fig. 8, it can be seen that when $x_{Ge}$ is around 0.3, the interdiffusivity in fully compressively strained SiGe is close to the B diffusivity in Si and in $Si_{0.8}Ge_{0.2}$, which is relevant to npn HBTs. In pMOS application, the source/drain stressors are B-doped, and are normally partially compressively strained. Assuming 50% of the compressive strain is relaxed, when $x_{Ge}$ is around 0.5, the interdiffusivity will be faster than the B diffusivity in Si and in $Si_{0.8}Ge_{0.2}$.

The reason that Si-Ge interdiffusion is not so obvious lies in the fact that we consider interdiffusion and dopant diffusion in different scales. When considering the interdiffusion of Ge profiles, we commonly use a linear scale. For example, in HBTs, one atomic percent (1 at.%) Ge peak drop is common after thermal activation, which is a concentration change of $5 \times 10^{20}$ cm$^{-3}$. This is considered as a small change in the linear scale. However, for dopant diffusion, a semilog scale is normally used, and a change of $1 \times 10^{19}$ cm$^{-3}$ (0.1 at.%) is considered to be quite significant.

Interdiffusion degrades MOSFET performance by reducing strain and carrier confinement and by increasing alloy scattering [3, 12]. In HBTs, the change of Ge distribution impacts the band engineering and material properties, which needs to be carefully designed and closely monitored. For both MOSFET and HBT industry, higher and higher Ge contents are being used. For Ge-based photonic devices, interdiffusion changes Ge to SiGe alloys, which is more indirect bandgap then Ge and undesired. It decreases photodetector efficiency [13] and reduces the direct band transitions and photoluminescence intensity, which delays the lasing of Ge/Si lasers [7].

This review starts with basic properties and the applications of major group IV semiconductors, and then reviews the progress made so far on Si-Ge and Ge-Sn interdiffusion behaviors. Theories, experimental methods, design and practical considerations will be discussed together with the key findings in this field.

### 1.2 Basic properties

Ge is the most Si-compatible semiconductor because it has the same crystalline structure (diamond cubic structure), a close lattice constant (a 4.18% larger lattice constant) and similar properties such as the energy band structure, stiffness constants, Poisson ratio, self-diffusivity, and specific heat. Ge and Si are miscible in the full concentration range. They are both indirect bandgap semiconductors. At 300 K, the indirect bandgap energy of Si is 1.12 eV. The direct and indirect bandgap energies of Ge are 0.80 and 0.66 eV respectively at 300 K. Ge can be engineered to become a pseudo-direct or direct bandgap semiconductor by the introduction of tensile strains and/or n-type doping [18]. Tensile strains lower Ge's direct energy bandgap energy faster than the indirect bandgap energy. Large enough tensile strain (biaxial tensile strains > 1.8% or uniaxial tensile strains > 4.6%) can convert Ge to a true direct-bandgap semiconductor without any doping [14-18].

Similarly, the energy band structures of $Ge_{1-x}Sn_x$ alloys can be engineered by the Sn content. Low Sn content GeSn alloys are of particular interests for applications in MOSFETs due to their high hole mobilities. They can also be made as light emitting diodes, photodetectors and lasers due to the direct bandgap suitable for mid-infra-red absorption and emission [19-21]. The introduction of Sn can lower Ge's direct energy bandgap faster than the indirect bandgap. Based on experiments and their extrapolation, the indirect to direct band transition of $Ge_{1-x}Sn_x$ is at around $x = 0.09$ for relaxed $Ge_{1-x}Sn_x$. On the solubility aspect, Ge and Sn form a binary eutectic alloy, with the eutectic temperature of 231.1 °C. The thermal equilibrium solid solubility of Sn in the Ge matrix is as low as 1 at.% below 500 °C [22]. As a result, it is difficult to increase the Sn content in $Ge_{1-x}Sn_x$ alloys because Sn precipitates from solid $Ge_{1-x}Sn_x$ easily during the crystal growth and post-growth processes. In recent years, Sn incorporation above the solid solubility has been achieved using chemical vapour deposition (CVD), molecular beam epitaxy (MBE) and solid state-epitaxy methods, as summarized in [22, 23].

Less than 0.5 at.% carbon in Si or SiGe is mainly used in SiGe HBTs for its capability in defect and thus diffusion and junction engineering. The current understanding is that carbon diffuses via an interstitial mechanism, which reduces the concentration of Si self-interstitials [24-26]. This in turn retards the diffusion of interstitial diffusers such as boron in Si and $Si_{1-x}Ge_x$ ($x < 0.2$) and P in Si. This approach is especially useful for NPN SiGe HBTs, which has been utilized in the industry since late 80's. The carbon incorporation approach is less effective for P in PNP SiGe HBTs, due to the small percentage of vacancy-assisted diffusion for P in $Si_{1-x}Ge_x$ ($x < 0.18$).

## 1.3 Applications

In the past three decades, due to the compatibility with the mainstream Si processing and the capability of energy bandgap/band structure, mobility, strain, and diffusion engineering, SiGe, SiGe:C, and Ge have been commercially and widely used in electronic and optoelectronic devices (Fig. 1) such as MOSFETs [27-30], tunnel field-effect transistors (TFETs) [31,32], SiGe:C hetero-junction bipolar transistors (HBTs) [8,9,33-36], photodetectors [37-39], modulators [40-42], and waveguides [43-45]. Besides these industry applications, many SiGe devices have been under extensive research such as SiGe nanowires transistors [46], Si and Ge nanocrystals thin film transistors and light emitters [47,48], Ge quantum dots lasers [49], Ge quantum dot photodetectors [50], GeSn lasers [21, 51], Ge lasers [52-59], SiGeSn light emitting diodes (LEDs) and photodetectors [58, 59].

The advantages of using SiGe, Ge, Si:C, GeSn are multifold. First, as discussed above, SiGe, Ge and GeSn have higher hole mobility than Si, which can be used as high hole mobility channel materials to boost MOSFET performance. Secondly, the lattice-mismatch can be used to introduce tensile or compressive strain, which changes energy band structures and carrier mobilities as well. These two approaches can be combined together to enhance carrier mobilities. Si:C can serve as tensile stressors for Si and SiGe channel nMOS. SiGe and GeSn can serve as compressive stressors for SiGe or Ge channel pMOS with smaller lattice constants. Thirdly, these materials can be used to engineer the bandgap structures such as in SiGe HBTs and SiGe TFETs. Fourthly, carbon is used to engineer point defect densities and thus junctions. Overall, the addition of the other elements gives more dimensions in engineering the material properties of the host material.

### 1.3.1 SiGe and GeSn in FETs

Si-based MOSFETs have been the most important and widely used semiconductor transistors, which are the fundamental building blocks of digital ICs. Since the 90 nm node production in 2004, SiGe has been used commercially in the state-of-the-art digital integrated circuits as source and drain stressors in p-type MOSFETs to introduce compressive stress to the channels and enhance hole mobility including the recent the 10 nm node FinFETs technologies. In a 2018 report by MSSCORPS Corporation Ltd., cross-section electron microscopy (XTEM) images and energy dispersive X-ray spectroscopy (EDS) mapping images of two cell processors, Exynos8895 and A11 Bionic used in Samsung's Galaxy S8 and iPhone 8 respectively, are shown. SiGe stressors for pMOSFETs are clearly seen in these images. SiGe-channel FinFETs with 25% Ge were used as 10 nm node pFET channels [60]. The benefits include better current drive and reliability [60]. 40% Ge channel FinFETs were reported in 2017 as seen in Fig. 9a [61].

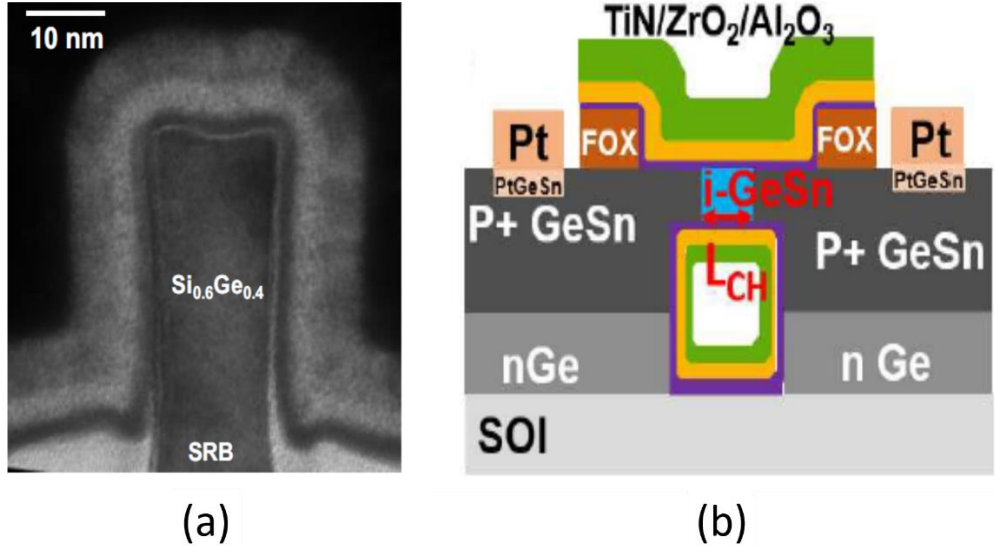

Fig. 9 SiGe and GeSn applications in transistors: (a) TEM image of a compressively-strained $Si_{0.6}Ge_{0.4}$ pFinFETs on a strained relaxed buffer (SRB). (b) Schematic cross-sectional view of single-GeSn channel MOSFETs. Figure reproduced from Refs. [61, 62] with permission from IEEE.

There are various GeSn electronic device applications, such as high-mobility $Ge_{1-x}Sn_x$ channel MOSFETs with up to 10% Sn as in Fig. 9b [62], high mobility channel MOSFETs with $Ge_{1-x}Sn_x$ source and drain stressors [63], and tunnel field-effect transistors (TFETs) [64] for high-performance and low-power consumption devices.

**1.3.2 SiGe:C in HBTs**

HBTs were dominated by III-V compound semiconductors before the emergence of SiGe HBTs, which were enabled by the lattice-matched strained-SiGe epitaxy technology. The use of SiGe:C in HBTs can be dated back to the late 80's, when it was first demonstrated by IBM [33]. SiGe is used to form a base with a narrower bandgap. Ge composition is typically graded across the base to create an accelerating electric field to help minority carriers to move faster across the base. A peak Ge molar fraction of 20% is commonly used in commercial products, and 30% Ge has been investigated in R&D.

For HBTs in high-speed and high-frequency applications, there are two important figures-of-merit, which are the common-emitter short-circuit cut-off frequency $f_T$ and the maximum frequency of oscillation $f_{max}$. Higher $f_T$ and $f_{max}$ represent better HBT device performance. Reducing the base width is an effective way to reduce the base transit time and thus increase $f_T$. Physically, higher base doping reduces electrical resistance, while narrower base doping profile reduces carrier transit time in the base region. These all contribute to a higher $f_T/f_{max}$. Heinemann et al. [65] recently reported experimental results of $f_T/f_{max}$ for up to 505/720 GHz. Aggressively scaled base doping profiles and the use of advanced annealing techniques are among the most important approaches to improve $f_T/f_{max}$. Carbon incorporation is one of the most important approaches to control B diffusion in NPN SiGe HBTs, creating narrower bases. The mechanism lies in the fact that C reduces interstitial concentration, which in turn reduces B diffusivity in Si and SiGe as B diffuses 100% by interstitials [66]. Carbon molar fractions used are generally from 0.1% to 0.3%. In SiGe:C systems, carbon is normally treated as an impurity instead of an alloy component. The major purpose of carbon incorporation is for point defect engineering. Due to the small concentration of carbon and the low solubility in SiGe, the focus of carbon in SiGe is on its substitutional portion.

For SiGe NPN HBTs and PNP HBTs, the base layers are commonly a few nanometers wide and are inside triangle-shaped Ge profiles. The main stream annealing is done by rapid thermal annealing (RTA) with a peak temperature in 1000 to 1100 °C range. Changes in the Ge profiles now is about 1%–3% at the triangle tip part of the Ge profiles. With the scaling in width, the increase in the Ge concentration and compressive strain, more Ge interdiffusion is expected, which has a big impact in the energy band engineering, the processing condition design and the performance of SiGe HBTs.

### 1.3.3 Ge and GeSn for Si compatible lasers

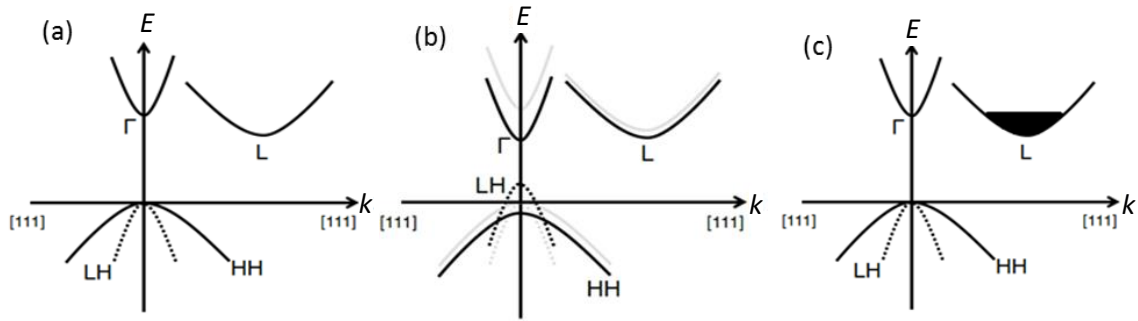

Fig. 10. Schematic diagrams of band structures of Ge under different conditions. (a) Ge without strain or doping; (b) tensile strained Ge without doping; (c) highly n-type doped Ge without strain. Figure reproduced from Ref. [52].

Ge and GeSn have also been studied for Si-compatible lasers [21, 51-57, 67-73] in the past two decades. The major motivation to study group IV lasers is their material and processing compatibility with the Si manufacturing infrastructure. Another important factor is in that Ge can be engineered into a pseudo-direct bandgap or direct bandgap semiconductor by adding tensile stresses or n-type doping or Sn as its direct bandgap is only 136 meV larger than its indirect bandgap.

In 2007, Liu et al. [52] from MIT showed that Ge can become a pseudo-direct bandgap material by adding tensile strain and/or high n-type doping as shown in Fig. 10. The n-type doing is used to fill the bottom energy states of the L valley such that the energy of the remaining available states can be higher than the bottom of the Γ valley of the direct band. In 2010, an optically pumped Ge-on-Si laser was first realized by adding a 0.24% biaxial tensile strain [53]. It operated with a gain of 50 cm$^{-1}$ at an n-type doping of $1 \times 10^{19}$ cm$^{-3}$ at room temperature. The lasing wavelength range was from 1590 to 1610 nm. In 2012, an electrically pumped Ge-on-Si laser was first demonstrated by researchers from MIT and APIC Corporation. It worked at room temperature and had a P doping level of $4 \times 10^{19}$ cm$^{-3}$ and a 0.2% biaxial tensile strain, as shown in Fig. 10c. The lasing wavelengths were from 1520 to 1700 nm due to different clamping conditions and the output power was up to 7 mW at room temperature [54]. The threshold current density was 280 kW cm$^{-2}$.

Tensile strains can be introduced by thermal expansion coefficients mismatch between Si substrates and Ge or by stress concentration methods. As shown in Fig. 10b, with an additional tensile strain applied to Ge, the Γ valley in Ge shrinks faster than the L valley due to the smaller effective mass in the Γ valley. For Ge to become a direct band gap material, according to calculations and experiments, the applied tensile strains need to be > 1.8% for biaxial strains or > 4.6% for uniaxial strains [15]. However, such high tensile strains narrow the bandgap too much so that the lasing wavelength will be larger than two microns [74]. Besides, it is technically hard to achieve such high tensile strain in Ge. There have been successful experimental efforts to change Ge into a direct bandgap material by introducing high uniaxial tensile strain up to 5.7% [74]. However, up to now, no successful efforts have been reported in making undoped Ge-on-Si lasers with tensile strain only.

Si-Ge interdiffusion has been shown to be mediated mostly by vacancies [5, 75, 76]. For Ge-on-Si lasers, due to an n-doping-introduced-increase in vacancy concentrations, interdiffusion during annealing steps has been shown to increase by 1 to 7 times for mid-$10^{18}$ cm$^{-3}$ P-doped Ge on Si [5,6]. Defect annealing steps are commonly needed to reduce TDD for Ge epitaxy. In our recent work, photoluminescence (PL) intensity measured from n-doped Ge-on-Si after defect annealing is much reduced from that of unannealed samples, and more indirect band transitions are observed after annealing [6]. This shows the competing effects between the defect reduction and interdiffusion. The PL data suggested that the negative impact from interdiffusion is larger than the benefits from the defect annealing. Therefore, a defect annealing step may not be needed for n-doped Ge/Si.

GeSn lasers emerged after Ge-on-Si lasers. S. Wirth reported the first optically pumped GeSn laser in 2015 with 12.6% Sn [71]. The working temperatures were below 90 K, and the threshold current density was 325 kW cm$^{-2}$. Later reported GeSn lasers showed operations up to 130 K and the wavelengths were between 2 to 3.1 µm [72, 73].

### 1.3.4 Ge and GeSn as photodetectors and waveguides

As discussed above, Ge has an indirect bandgap 0.66 eV at room temperature, which corresponds to 1.55 µm in wavelength. As a result, its optical applications are mainly in short infrared or mid-infrared region. Light detection and modulation can be realized with Ge-on-Si waveguides. Ge waveguide p-i-n photodetectors have achieved 67 Gb s$^{-1}$ operation [77]. Ge-on-Si waveguides can work for 1.5486 µm wavelength [78] and larger wavelength, such as between 5.2 and 5.4 µm with a minimum propagation loss of 3 dB cm$^{-1}$ [79]. A more recent experiment had operating wavelengths between 7.5 and 8.5 µm and a minimum propagation loss of 2.5 dB cm$^{-1}$ at 7.575 µm wavelength [45].

The indirect and direct bandgap of Ge$_{1-x}$Sn$_x$ alloys decreases with the Sn concentration. Therefore, the corresponding wavelengths are larger compared to Ge photodetectors and waveguides. The transition to a direct bandgap GeSn happens at around 8 at.% Sn for relaxed GeSn. Theoretically, the wavelength could be extended to 2.32, 2.69, and 4.06 µm for $Ge_{0.97}Sn_{0.03}$, $Ge_{0.95}Sn_{0.05}$, and $Ge_{0.90}Sn_{0.10}$ waveguide photodetectors under 1% tensile strain respectively [58]. Photodetection up to 2.2 µm was achieved experimentally by $Ge_{0.91}Sn_{0.09}$ quantum wells on silicon substrate [80]. Tseng et al. [81] reported GeSn photodetectors with a larger photo-responsivity with only 2% Sn concentration. The operation speed can reach 6.2 GHz. Quantum efficiency analysis indicated that Sn introduction can also reduce the device length to achieve the maximum efficiency about 60% at 1550 nm wavelength in waveguide photodetectors [82].

Preliminary GeSn lasers work at very low temperatures such as 90 to 130 K was discussed previously. Amplified spontaneous emission was observed by optically pumped GeSn waveguides at room temperature [83]. However, the net cavity gain could not exceed the lasing threshold in the experiment, which is a barrier for GeSn laser development [59].

Another important application for GeSn alloys is in light-emitting diodes (LED), which are more mature than GeSn lasers. GeSn LEDs with an indirect bandgap have been well studied at temperatures from 77 to 300 K with Sn concentrations up to 9.2% [84]. Direct bandgap GeSn LEDs have been demonstrated, which have Sn concentration up to 11% and the emitted phonon energy is 0.55 eV (wavelength = 2.25 µm) at room temperature [21].

## 2. Research approaches in interdiffusion studies and the benchmarking Si-Ge interdiffusivity

### 2.1 Fundamental theories and concepts

Before we discuss further, some fundamental theories need to be reviewed for the convenience of the readers. These can be found in many classic books on diffusion such as Ref. [85].

In most (inter)diffusion cases relevant to semiconductor processing, chemical potential gradients are due to chemical composition gradients, and most (inter)diffusion cases in semiconductor processing are Fickian diffusion, where Fick's Laws apply. Fick's first law relates the chemical flux of a species $F$ with its chemical composition C gradient in the crystal frame (C-frame) as

$$F = -D\frac{dC}{dx}. \qquad (3).$$

Here, $D$ is the intrinsic diffusivity of this species in a specific diffusion medium.

From thermodynamics and basic diffusion theories, intrinsic diffusivity $D$ of element A can be expressed as:

$$D_A = D_A^* \left(\frac{\partial \ln a_A}{\partial \ln x_A}\right). \qquad (4)$$

In the above equation, $D_A^*$ is the self-diffusivity of A, which is the diffusivity of one isotope of A in chemically homogenous solid solutions. Self-diffusion is only possible for a mixture of isotopes where there is no chemical gradient and the thermal motion of different isotopes can be monitored. $a_A$ and $x_A$ denote the chemical activity and the molar fraction of component A. The term $\frac{\partial \ln a_A}{\partial \ln x_A}$ counts for the chemical mixing effects. For ideal solutions, such as mixtures of different isotopes of one element, $\frac{\partial \ln a_A}{\partial \ln x_A} = 1$.

The self-diffusivity $D^*$ can be macroscopically expressed as

$$D_A^* = MRT, \qquad (5)$$

where $M$, R, and $T$ denote the mobility of element A, the ideal gas constant, and the absolute temperature.

In reality, semiconductors with low to medium doping levels, commonly less than 1 at.%, are very close to ideal solutions. SiGe alloys, on the other hand, have comparable Si and Ge concentrations, and thus cannot be treated as ideal solutions. For non-ideal solutions, $\frac{\partial \ln a_A}{\partial \ln x_A} \neq 1$, and the influence from chemical mixing cannot be ignored. In these cases, the diffusivity $D$ is intrinsic diffusivity. The relation of self-diffusivity and intrinsic diffusivity can be expressed as

$$D_A = D_A^* \left(\frac{\partial \ln a_A}{\partial \ln x_A}\right) = D_A^* \left(1 + \frac{\partial \ln \gamma_A}{\partial \ln x_A}\right), \qquad (6)$$

where $\gamma_A$ stands for the activity coefficient of A.

For a binary alloy system with element A and B, in a laboratory coordination system (L-frame), the proportional factor $\widetilde{D}$ is the interdiffusivity.

$$F = -\widetilde{D} \frac{dC}{dx}. \qquad (7)$$

Some may ask about the differences between the intrinsic diffusivities of A and B in the alloy and the A-B interdiffusivity. These parameters are not the same because they refer to two different reference coordination systems known as "frames". The C-frame may change during the interdiffusion. For example, in a Si/Ge interdiffusion couple, if a marker layer is placed at the interface, it moves during the interdiffusion as one side loses its lattice sites and the other side gains more lattice sites. The changes in lattice site numbers are by the movement of vacancies during interdiffusion. Vacancies are missing lattice sites, which are one major type of point defects. The side that receives the net vacancy flux will shrink, while the volume of the other side increases. This means that experimentally, the marker layer moves towards the side that loses its lattice sites, which is a net movement in reference with the lab frame (L-frame). This is called "Kirkendall effect".

The Fickian diffusion of each species in the C-frame, a coordinate system with respect to the starting interface or a marker layer pinned to the starting interface, follows Fick's first law, with a proportionality constant known as the intrinsic diffusivity.

The C-frame is not the L-frame, which is a volume-fixed frame. Therefore, the intrinsic diffusivity D of A or B in the alloy is used in the C-frame, while interdiffusivity $\widetilde{D}$ is used in the L-frame, which is also the common frame that we use in the experiments without marker layers pinned to the initial interfaces. Darken's law was first reported by Darken in 1948 for binary metallic systems [86]. It describes the correlation between self-diffusivity, intrinsic diffusivity and interdiffusivity with equilibrium point defects. Intrinsic diffusivities of A and B and the A-B interdiffusivity are related by the Darken equation below [86]:

$$\widetilde{D} = D_A x_B + D_B x_A. \qquad (8)$$

According to Darken's law, at the B end, where $x_A = 0$ and $x_B = 1$, $\widetilde{D} = D_A$, while at the A end, $\widetilde{D} = D_B$. In the two special cases, the interdiffusivity is the intrinsic diffusivity of A or B.

Combining Eqs. (6) and (8), the interdiffusivity is

$$\widetilde{D} = D_A^* \left(1 + \frac{\partial \ln \gamma_A}{\partial \ln x_A}\right) x_B + D_B^* \left(1 + \frac{\partial \ln \gamma_B}{\partial \ln x_B}\right) x_A. \qquad (9)$$

If the self-diffusivities and the dependence of $\gamma$ on $x_{Ge}$ are known for both A and B, we can then calculate the intrinsic diffusivities and the interdiffusivity as a function of $x_{Ge}$. This is exactly the idea of the self-diffusivity approach, which is discussed in Session 2.5.

## 2.2 Background of Si-Ge interdiffusion study and practical considerations

Major Si-Ge interdiffusion research efforts started in the 1990's. Several groups measured interdiffusion with various techniques such as SIMS [87], Rutherford backscattering spectrometry [88,89], X-ray diffraction (XRD) [90,91], Raman spectroscopy [92] and photoluminescence [93]. Typical interdiffusion structures studied in the 90's were Si/Si$_{1-x}$Ge$_x$ superlattices with thickness from 30 nm to a few microns and in low Ge ranges ($x_{Ge} < 0.3$). Due to device scaling and advancement in structures and fabrication techniques, typical Si-Ge interdiffusion lengths in current technologies are in 1 to 100 nm range, comparable to the thickness of SiGe thin films in the devices. $x_{Ge}$ range of current interest is now much higher than previous studies. Therefore, in the past one to two decades, Si-Ge interdiffusion has been revisited.

The complications of binary semiconductor alloy systems lie in the concentration, strain, defect density, point defect engineering (such as carbon incorporation and oxidation) and doping, which are all technologically significant and they impact the interdiffusion simultaneously. For example, Si-Ge interdiffusivity has a very strong dependence on the Ge concentration. For relaxed SiGe with low defect density and no intentional doping in the temperature range from 700 to 1000 °C, the interdiffusivity at the Ge end is 5 to 6 orders of magnitude larger than that at the Si end [94]. It increases exponentially with compressive strains, and increases with n-type dopants and implant damages [1, 4, 6, 94, 95].

To be relevant to the modern industry practice, atomic-scale interdiffusion studies require a high material quality, a careful experiment design to separate the impacting factors, and a high accuracy in the thermal annealing temperatures and characterization tools. Here are some key experimental considerations.
(1) High quality epitaxial systems with low dislocation density (commonly below mid-$10^{16}$ cm$^{-3}$ and well-defined pre-interdiffusion concentration profiles are required.
(2) Strain status needs to be closely monitored such that interdiffusion can happen with a known strain history. This in turn requires a careful design of the film stacks and thermal budgets.
(3) Temperature calibration is crucial to the accuracy of data and modeling. A temperature accuracy of a few degrees is required. Mainstream rapid thermal anneal (RTA) tools work by thermal absorption from intense light emissions. RTA tools are very sensitive to the chamber cleanness, and are much harder to calibrate for general purpose cleanrooms, where various materials are used. A practical solution is to use industry R&D facility where the annealing tools are well-maintained and calibrated for a narrow group of material systems. However, industry tools are normally hard to get access to. Another option is to use a heating stage or a furnace with resistive heating.
(4) Atomic scale Si and Ge depth profiling are normally done by secondary ion mass spectrometry (SIMS). Although SIMS tools are not rare in Canadian and US schools, we have not found any SIMS tools in an academic setting that is calibrated enough to cover the full Ge fraction ranges with sub-nm depth resolutions. For our studies at MIT and the University of British Columbia (UBC), SIMS measurements have been commercially performed by Evans Analytical Group, who provides good data repeatability and accuracy.

In the discussions below, we will mostly use the Si-Ge binary system as an example to discuss the methods used in interdiffusion studies. Generally there are three major approaches to study Si-Ge interdiffusion experimentally.

## 2.3 XRD approach

The first approach uses high resolution X-ray diffraction (HRXRD) to probe multiple layered superlattice structures, which can also be considered as multiple quantum wells (MQWs). This approach was used by researchers such as Aubertine, Meduna, Ozguven and Liu et al. [96-100]. The principle underlying this technique is the correlation between the interdiffusivity and Bragg reflection intensity decay rate of superlattice satellite:

$$\frac{d}{dt}\left[\frac{I(t)}{I(0)}\right] = -\frac{8\pi^2}{\lambda^2}\widetilde{D}, \tag{10}$$

where $I(t)$ is the superlattice satellite intensity measured by XRD as a function of the annealing time $t$, $I(0)$ is the original intensity at $t = 0$, $\lambda$ is the spatial period of the superlattice, and $\widetilde{D}$ is the interdiffusivity.

This technique utilizes the ultrahigh sensitivity of XRD for concentration modulated films. However, this method only solves one interdiffusivity for one superlattice, which is an averaged interdiffusivity for a concentration range. As Aubertine et al. [96] pointed out, when it was applied to concentration dependent interdiffusion like Si-Ge interdiffusion, this method ignores the strong dependence of the interdiffusivity on $x_{Ge}$, and is only suitable for superlattices Si$_{1-x}$Ge$_x$/Si$_{1-y}$Ge$_y$ with a small difference in $x$ and $y$. This method also ignores the strain distribution in a superlattice, which has a tensile/compressive pattern. The limitation of this method is that it is an indirect way of monitoring interdiffusion, compared to SIMS. For each Ge fraction studied, a superlattice structure with a narrow Ge fraction amplitude needs to be grown, which is normally not device-related structures. The advantage of this method is that XRD is a very sensitive non-destructive method, which can monitor the time evolution of interdiffusivity. It is suitable for a lab with a good epitaxy growth capability.

Aubertine et al.'s [96] work focused on the interdiffusion in MQW with $x_{Ge}$ below 0.2 and in a temperature range from 770 to 880 °C. Figure 11 shows the schematic as-grown and annealed profiles. 113 Bragg reflection was used to measure the films' average Ge concentration and the strain status. The interdiffusion was modelled with a concentration-dependent diffusivity pre-factor $D_0$ and activation energy $E_a$. Ozguven et al. [97] measured the interdiffusivity at $x_{Ge, \text{average}}$ = 0.91. To avoid too much averaging over a large Ge range, the Ge molar fraction difference in each MQW is 0.05.

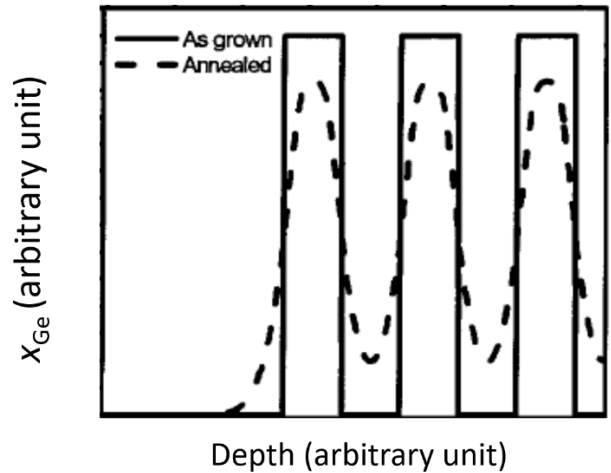

**Fig. 11. Schematic as-grown Ge molar fraction profile and annealed Ge profile of a multiple quantum well structure. Figure after Ref. [101] with permission from AIP Publishing.**

With much larger Ge molar fraction differences from 0.15 to 0.80 in one period of MQWs, Meduna et al. [98, 99] investigated four $x_{Ge, \text{average}}$ (from 0.25 to 0.90) and obtained similar interdiffusivity formulas to Aubertine's. Liu et al. [100] measured it at $x_{Ge, \text{average}}$ = 0.85 with a $x_{Ge}$ difference of 0.35 in one period of MQWs. There are two significant issues with large $x_{Ge}$ differences in one period of MQWs. First, the $x_{Ge}$ corresponding to the interdiffusivity is taken as the $x_{Ge, \text{average}}$ of MQWs or measured by Rutherford backscattering spectrometry. This is not a good method to obtain the nominal Ge value as the interdiffusivity depends on $x_{Ge}$ exponentially instead of linearly. According to the model reported by Gavelle et al. [101], for one period of $Si_{0.45}Ge_{0.55}$/Ge MQW, the interdiffusivity in the compressive Ge layer is almost 1000 times larger than that in the $Si_{0.45}Ge_{0.55}$ layer. Once the interdiffusion starts, Ge from the higher Ge regions diffuses out to the lower Ge molar fraction layer, which reduces the difference.

Recently, interdiffusion in a Ge-Sn system was studied using an XRD approach [102]. Ge/$Ge_{0.9}Sn_{0.1}$ MQW structures were used. The annealing temperatures studied were from 300 to 600 °C. For the temperature range from 380 to 450°C, the effective interdiffusivity is in the range of $10^{-16}$ to $10^{-15}$ cm²/s, where the Sn molar fraction $x_{Sn}$ ranges from 0 to 0.1. They described the interdiffusivity using an Arrhenius relation.

$$\widetilde{D}_{\text{effective}} = 2.9 \times 10^{-7} \text{cm}^2\text{s}^{-1} \exp(-\frac{1.21\text{eV}}{kT}) \qquad (11)$$

So far, this work has been the only experimental work that we are aware of on Ge-Sn interdiffusion. More detailed studies on Ge-Sn interdiffusion need to be performed in the future.

## 2.4 SIMS approach

The second approach employs SIMS. Cowern et al.'s work in 1994 used SIMS to measure diffused Ge profiles, and used diffusion models to fit the profiles [87]. Xia et al.'s [1, 103] work in 2006 and 2007 first introduced Boltzmann-Matano method to extract Si-Ge interdiffusivity $\widetilde{D}_{\text{Si-Ge}}$ as a function of $x_{Ge}$ from the diffused profiles of step structures, as shown in Fig. 12a. Boltzmann-Matano method can be used to extract the interdiffusivity from an annealed concentration profile when the pre-diffusion profile can be approximated as a step profile, also known as an interdiffusion couple. This method also requires that the interdiffusivity is only dependent on the alloy concentration. This condition can be satisfied when an epitaxial system is pseudomorphic, where the strain dependence can be expressed as a Ge concentration dependence and the annealing is isothermal with no significant change in the strain or the dislocation density. The isothermal condition the interdiffusion couple condition is easy to satisfy. However, other conditions are not easy to satisfy. For example, if there is strain relaxation or a non-uniform distribution of threading dislocations or doping, the interdiffusivity is no longer a function of the Ge concentration only. These scenarios are common in the industry practice, in which cases, we can still use this method to estimate the interdiffusivity, which is then called the effective interdiffusivity.

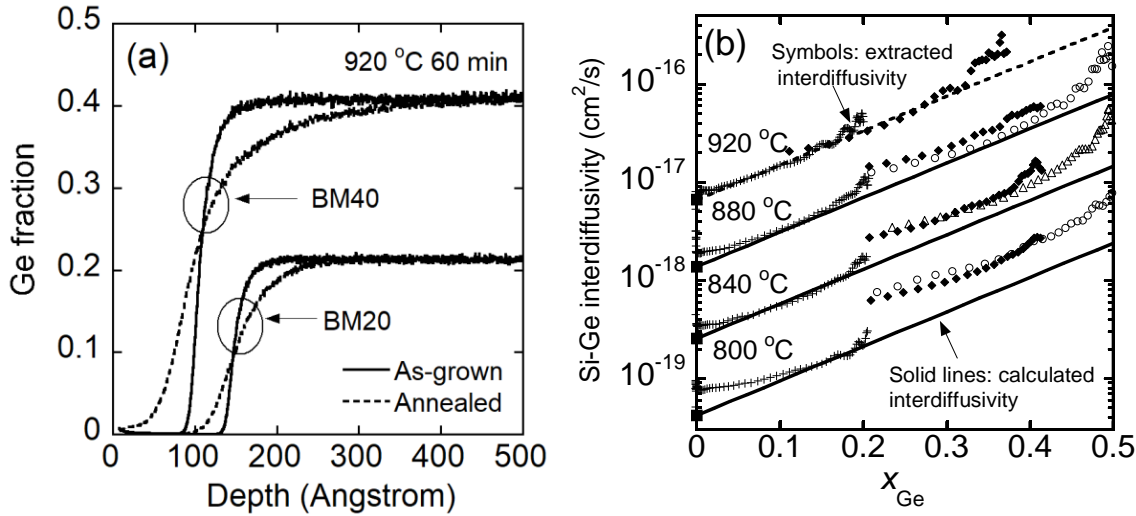

**Fig. 12. Example of the application of Boltzmann-Matano method in Si-Ge interdiffusivity extraction: (a) Ge profiles of the interdiffusion couples up to 40% Ge before and after diffusion. (b) Interdiffusivity extracted from Boltzmann-Matano analysis (symbols) compared to the $D_R$ model expressed in Equation 12 (solid lines). Squares at $x_{Ge} = 0$ are Ge tracer diffusivity data. Figure adapted from Ref. [1] with permission from AIP Publishing.**

Compared with the XRD approach, the SIMS and Boltzmann-Matano analysis approach makes it convenient to extract $\widetilde{D}$ as a function of $x_{Ge}$. It also works well to include the strain impact, as it can be expressed as a function of $x_{Ge}$ in coherent epitaxial films. The step structures are easier to grow than superlattice structures. Most importantly, the diffusion structures are device-relevant.

Using this approach, Xia et al. [1] built a $D_R D_C$ model to describe the Si-Ge interdiffusivity under relaxed, tensile and compressive stress as shown in Fig. 12 (b). This work reveals the exponential dependence of interdiffusivity on the Ge fraction and biaxial compressive strain.

$$\widetilde{D}_{R,T}(x_{Ge}) = 310 \exp\left(-\frac{4.66\ eV}{kT}\right) \exp(8.1 x_{Ge}), \tag{12}$$

$$\widetilde{D}_C(x_{Ge}) = \widetilde{D}_{R,T}(x_{Ge}) \exp\left(-s\frac{|\varepsilon_C|}{kT}\right), \tag{13}$$

where s is approximately 15, $\widetilde{D}_{R,T}$ refers to interdiffusivity under relaxed or tensile strain, and $\widetilde{D}_C$ refers to that under compressive strain. All strains here are biaxial due to planar growth of thin epitaxial films.

The error bar of this model is from 50% to 150% of the measured interdiffusivity, which is mainly from the annealing temperature uncertainty and SIMS errors. Although the $D_R D_C$ model lacks a detailed theoretical study on the activation energy dependence, it catches the interdiffusion behavior (the Ge fraction, stress and temperature dependence) with quite reasonable accuracy and an easy to implement format. In 800 to 1200 $^0$C, the temperature range relevant to technology, this model has been used to predict interdiffusion for both furnace anneals and RTAs, and has been shown to be reasonably accurate as seen in Dong et al.'s work [94]. Therefore, this model can be used as the first model to predict interdiffusion for $x_{Ge} < 0.56$.

With an amorphous Ge/Si structure, Si/Ge, Gavelle et al. [101] extracted and then modelled experimental interdiffusivity with a two-term equation. One term was for dislocation-mediated interdiffusion, and the other for point-defect-mediated interdiffusion. Due to the high dislocation density ($10^{10}$ cm$^{-2}$) in their structures, Gavelle et al.'s data are good to show the dislocation-mediated interdiffusion term. In the high Ge region, the data agree with Dong et al.'s model quite well, where the point-defect-mediated term dominates.

## 2.5 The self-diffusivity approach

Besides the two approaches discussed above, Dong et al.'s work at UBC first introduced the third approach in studying Si-Ge interdiffusivity under no stress or tensile stress, which is the self-diffusivity approach [94]. This approach is based on Si, Ge self-diffusivity and chemical activity coefficients. Interdiffusivity is calculated from self-diffusivity data instead of from the measurements of interdiffusion. The advantage of this approach is that it is able to cover a wide range of temperature and Ge fraction range, and is especially useful for interdiffusion under rapid thermal annealing conditions, where the temperature ranges are large. In addition, this interdiffusivity model provides a zero strain, no doping, and low dislocation density reference for later studies of more impacting factors of Si-Ge interdiffusion. To use this model for structures with compressive strains, a compressive strain enhancement factor, such as that in the $D_R D_C$ model in Equation (13) or that in Eqs. (22) and (23) (discussed in Session 3.3) need to be added to account for that.

### 2.5.1 Si and Ge self-diffusivity data and modeling

Si and Ge self-diffusivity ($D_{Si}^*$, $D_{Ge}^*$) studies based on Si and Ge isotope diffusion have been reported since 1970's and the focus was on the self-diffusivity of Ge isotopes in Si$_{1-x}$Ge$_x$ alloys [104-109]. Due to the shorter lifetimes of Si isotopes, however, there was little systematic study of Si self-diffusivity in Si$_{1-x}$Ge$_x$ alloys until the studies by Kube et al. [108,109]. In their work, $D_{Si}^*$, and $D_{Ge}^*$ were measured at six $x_{Ge}$ values.

These data enabled the practice to establish a quantitative relation between $D_{Si}^*$, $D_{Ge}^*$, and $\widetilde{D}$. However, these studies only measured $D_{Si}^*$, and $D_{Ge}^*$ at six discrete $x_{Ge}$ values. Some data interpolation was needed. An Arrhenius relation in Equation (11) was used, where $D_{Si,0}^*$ and $D_{Ge,0}^*$ are the prefactors and $E_a$ is the activation energy. These three parameters are all $x_{Ge}$ dependent.

$$D_j^*(x_{Ge}) = D_{j,0}^*(x_{Ge}) \exp\left(\frac{-E_{a,j}(x_{Ge})}{kT}\right), j = \text{Si or Ge}. \quad (14)$$

This can be plugged in Equation (9), and the Si-Ge interdiffusivity is expressed as

$$\widetilde{D} = D_{Si,0}^*(x_{Ge}) \exp\left(\frac{-E_{a,Si}(x_{Ge})}{kT}\right)\left(1 + \frac{\partial ln\gamma_{Si}}{\partial lnx_{Si}}\right)x_{Ge}$$
$$+ D_{Ge,0}^*(x_{Ge}) \exp\left(\frac{-E_{a,Ge}(x_{Ge})}{kT}\right)\left(1 + \frac{\partial ln\gamma_{Ge}}{\partial lnx_{Ge}}\right)x_{Si}, \quad (15)$$

To obtain practical $D_{Si,0}^*$ and $D_{Ge,0}^*$ the full Ge range, we need to interpolate Kube et al.'s data. $D_{Si,0}^*$ and $D_{Ge,0}^*$ can be expressed as

$$D_{Si,0}^*(x_{Ge}) = \exp(a_0 + a_1 x_{Ge} - a_2 x_{Ge}^2) \text{ and } D_{Ge,0}^*(x_{Ge}) = \exp(b_0 + b_1 x_{Ge} - b_2 x_{Ge}^2), \quad (16)$$

where $a_0$ = 6.489, $a_1$ = 4.964, $a_2$ = 7.829; $b_0$ = 6.636, $b_1$ = 8.028, $b_2$ = 11.318. All these fitting parameters are dimensionless.

The activation energy $E_{a,j}$ of self-diffusivity follows a modified Vegard's law shown in Equation (17), which has a second order dependence on $x_{Ge}$ [109]. The parameters $Q(0)$, $Q(1)$ and $\Theta$ are all from Ref. [109]. For Si self-diffusivity, they are 4.76, 3.32 and 1.54 eV respectively. For Ge self-diffusivity, they are 3.83, 3.13 and 1.63 eV respectively. This means that the activation energy for Si self-diffusivity is between 4.76 and 3.32 eV depending on the Ge molar fraction, while that for Ge is between 3.83 and 3.13 eV.

$$E_{a,j}(x_{Ge}) = (1 - x_{Ge})Q_j(0) + x_{Ge}Q_j(1) + x_{Ge}(1 - x_{Ge})\Theta_j, \ j = \text{Si or Ge}. \quad (17)$$

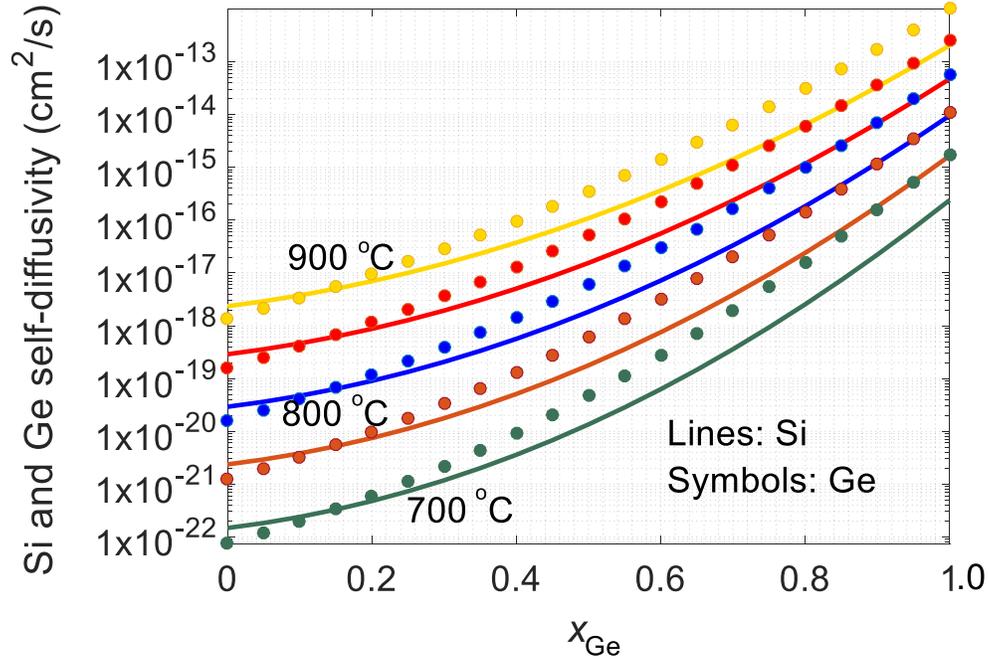

**Fig. 13. Calculated Si and Ge self-diffusivity values as a function of the Ge molar fraction in the temperature range from 700 to 900 °C based on Eqs. (14), (16) and (17).**

After $D_{Si}^*$, $D_{Ge}^*$ and $E_{a,j}$ were modeled, we calculated the Si and Ge self-diffusivities using the above equations. The calculated self-diffusivities show a good consistency with Kube et al. and Strohm et al.'s experimental data [106-108]. For the reader's convenience, Fig. 13 plots the Si self-diffusivity and Ge self-diffusivity using the above parameters and models. It can be seen that at a certain temperature, there are 5 to 6 orders of magnitude difference between the self-diffusivities at the Si end and those at the Ge end, showing a very strong dependence on the Ge molar fraction.

### 2.5.2 Interdiffusivity calculations

In the following derivations, SiGe solid solutions are assumed to be regular solutions, where the entropy of mixing is the same as that for an ideal solution. The partial molal enthalpy $\overline{\Delta H}$ of Si and Ge in a SiGe solid solution is shown in Equation (18) [110], where $\alpha$ is the interaction parameter. Bublik et al. [111] measured $\alpha$, which is linear with $x_{Ge}$ as in Equation (19). $\frac{\partial ln\gamma}{\partial lnx}$ for Si and Ge was calculated with Equation (20). Theoretically, $\frac{\partial ln\gamma}{\partial lnx}$ should be identical for Si and Ge. The calculated results based on the experimental $\alpha$ are quite close. The small difference between $\frac{\partial ln\gamma_{Si}}{\partial lnx_{Si}}$ and $\frac{\partial ln\gamma_{Ge}}{\partial lnx_{Ge}}$ is caused by the asymmetry of $\alpha$ over Ge molar fraction. After all these parameters are known, $\widetilde{D}_{Si-Ge}$ can be calculated using Eqs. (9), (14) to (20). We will refer to this calculation as Dong et al.'s model in the following.

$$\Delta \overline{H}_{Si} = RTln\gamma_{Si} = \alpha x_{Ge}^2, \text{ and } \Delta \overline{H}_{Ge} = RTln\gamma_{Ge} = \alpha x_{Si}^2 \quad (18)$$

$$\alpha = 8787 - 1339 x_{Ge} \text{ (J/mol)} \quad (19)$$

$$\frac{\partial ln\gamma_{Si}}{\partial lnx_{Si}} = \frac{(1-x_{Ge})x_{Ge}(4017x_{Ge}-17574)}{RT} \quad and \quad \frac{\partial ln\gamma_{Ge}}{\partial lnx_{Ge}} = \frac{(1-x_{Ge})x_{Ge}(4017x_{Ge}-18913)}{RT}, \qquad (20)$$

### 2.5.3 The benchmarking interdiffusivity

Dong et al.'s [94] model applies to interdiffusion under no strain and with tensile strain less than 1%. 1% tensile strain was shown to have little impact on the interdiffusion for SiGe with up to $x_{Ge}$ = 0.3. The tensile strain impact can also be seen from Fig. 12. The extracted interdiffusivities in Fig. 12b. are from different interdiffusion couples such as s-Si/r-Si$_{0.8}$Ge$_{0.2}$, s-Si/r-Si$_{0.6}$Ge$_{0.4}$, and s-Si/r-Si$_{0.44}$Ge$_{0.56}$. Here "s" means "strained", and "r" means "relaxed". The s-Si layers are under different tensile strains. If tensile strains have a significant impact, then the extracted interdiffusivity curves will not overlap or follow the same trend. Therefore, we conclude that tensile strain up to 1% has little impact. This was also confirmed in Session 3.3 discussed below.

The interdiffusivity at $T$ = 900 °C calculated using Dong et al.'s model is shown in Fig. 14, which agree with literature data and models very well. Dong et al.'s model also matches Gavelle et al.'s results well in $x_{Ge} \geqslant 0.85$ regime. Samples in Gavelle et al.'s work have a high dislocation density ($10^{10}$ cm$^{-2}$), while the samples in other literature work have dislocation densities in the $10^5$ to $10^6$ cm$^{-2}$ range. Therefore, Si-Ge interdiffusion in Gavelle et al.'s work has a large dislocation-mediated interdiffusion component. This term dominates and results in a much faster interdiffusion in the low to medium $x_{Ge}$ range. Besides the good agreement with literature data from multiple research groups, Dong et al.'s model gave very good predictions for Si/Si$_{1-x}$Ge$_x$ heterostructures under soak and spike rapid thermal annealing up to 1200 °C [94]. Its accuracy has been further confirmed with our later studies on strain and doping impacts when it served as the benchmarking interdiffusivity lines.

One thing to bear in mind is that Dong et al.'s model is based on the self-diffusivity data and their interpolation in Ref. [109]. The temperature range was from 880 to to 1270 °C at $x_{Ge}$ = 0 end (the Si end), and 550 to 900 °C at $x_{Ge}$ = 1 end (the Ge end) correspondingly. Therefore, it is valid in this temperature range. We recommend to use Dong et al.'s model in 800 to 1300 °C range for low Ge alloys, in 700 to 1100 °C range for mid-Ge range alloys and in 500 to 900 °C for high-Ge alloys.

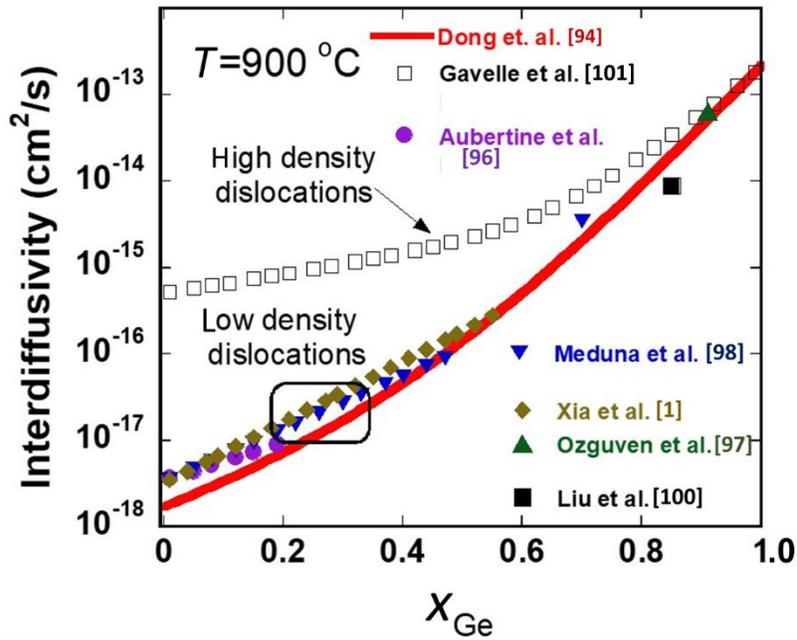

**Fig. 14.** SiGe interdiffusivity at 900°C calculated using Dong et al.'s model in comparison with literature data and models. Figure adapted from Ref. [94] with from AIP Publishing..

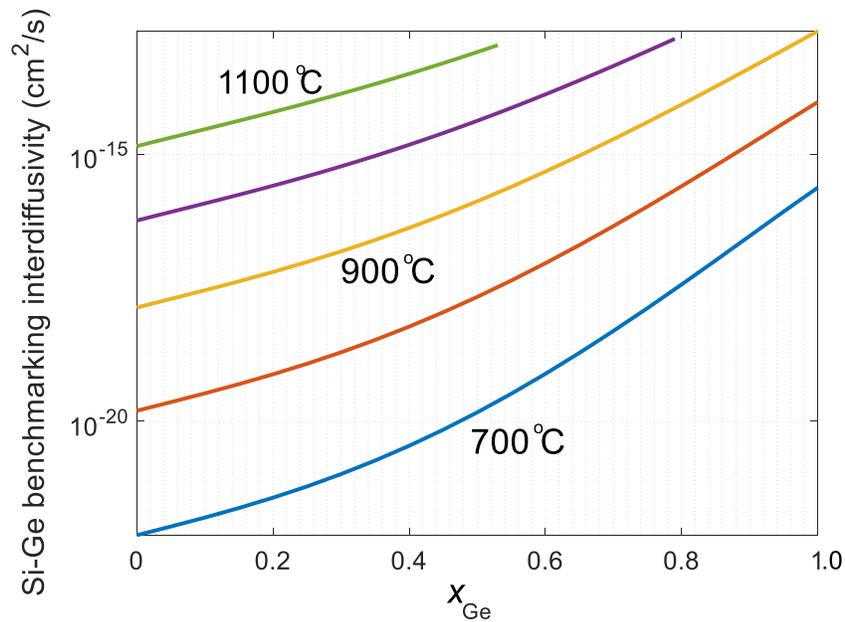

**Fig. 15.** Calculated Si-Ge benchmarking interdiffusivity based on Dong et al.' model in [94] for 700 to 1100 °C. These results apply to undoped, unstrained/tensile-strained SiGe systems with low defect densities (threading dislocation density < $10^7$ cm$^{-2}$). They served as the benchmarking lines for later interdiffusion studies on strain, defect and doping effects.

For the convenience of potential users, we provide more Si-Ge interdiffusivity curves in Fig. 15, which has the interdiffusivity results calculated for the temperature range from 700 to 1100 °C with an interval of 100 °C. Readers can use Eqs. (9), and (14) to (20) to calculate the interdiffusivity of the temperatures of their interests.

As the melting temperature of Ge is 938.8 °C, above this temperature, the Ge molar fraction range for solid-state SiGe alloys doesn't extend to 100% Ge. For example, according to the solidus line of the Si-Ge phase diagram under one atmosphere pressure, at 1000 °C, $Si_{1-x}Ge_x$ alloys are solid when $x < 0.79$. At 1100 °C, $Si_{1-x}Ge_x$ alloys are solid when $x < 0.53$. From Fig. 15, we can see that the interdiffusivity $\widetilde{D}$ has a very strong dependence on $x_{Ge}$. From $x_{Ge} = 0$ to 1, the interdiffusivity increases by 5 to 6 orders of magnitude in 700-900 °C for relaxed SiGe. The interdiffusivity $\widetilde{D}$ has a near-exponential relationship with $x_{Ge}$ at the Si end and at the Ge end. The $x_{Ge}$ dependence is stronger at the Ge end. Therefore, if the $x_{Ge}$ range is limited as the case in [1]'s study, the interdiffusivity can be approximated as an exponential relation with $x_{Ge}$. In terms of the temperature dependence, 8 to 17X enhancement is observed for every 50 °C increase at the Si end, and 4 to 7X enhancement for every 50 °C at the Ge end in 700 to 900 °C as seen in Fig. 16. These numbers are helpful to understand the temperature dependence, although there is not a simple activation energy for interdiffusion as expressed in Eq. (15).

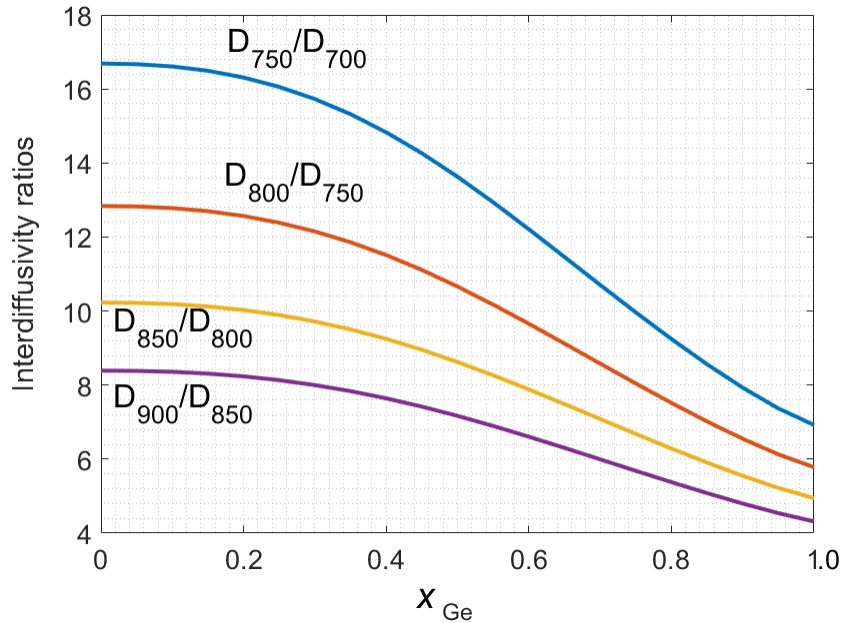

**Fig. 16. Interdiffusivity ratios showing the temperature dependence for undoped SiGe under tensile or relaxed strains. The calculations are based on Dong et al.' model in [94].**

# 3. Impacting Factors of Interdiffusion: Stress, Doping, Defect Engineering and Threading Dislocation Density

## 3.1 Background and terms

In the above benchmarking Si-Ge interdiffusivity $\widetilde{D}$, the temperature and Ge concentration are included. It is based on self-diffusivities, where tracer diffusion is measured in a chemically homogenous material with no strain, which is quite ideal. Stress, doping, defect engineering (such as oxidation and carbon incorporation) and defect density are all important impacting factors. These factors are very relevant to the industry practice. For example, in SiGe-channel p-type FinFETs shown in Fig. 9, the SiGe channel is under compressive strain, and defects such as dislocations are unavoidable. In source and drain stressors for CMOS applications, SiGe stressors are normally highly doped. In SiGe HBTs, carbon and dopant coexist in compressive SiGe base layers. In Ge-on-Si lasers, the Ge layers are commonly tensile-stressed with high n-type doping, and high defect density exists at the Ge/Si interfaces. Therefore, the studies on stress, doping, defect engineering and defect density are crucial to the industry practice.

Here, we need to clarify the term "defect engineering". "Defect" is a generic term, which include defects of 0D, 1D, 2D and 3D forms. When using the term "defect engineering" in the context of diffusion in semiconductors, point defects are commonly referred, which can be engineered using oxidation, nitridation and carbon incorporation to tweak the point-defect-mediated diffusion and thus dopant or alloy concentration distributions.

Dislocations are line defects, which add a dislocation-mediated term on top of the point-defect-mediated term. The modeling shown above in Equation (7) to (12) has no dislocation-mediated term, as it refers to SiGe material systems with low threading dislocation densities ($< 10^7$ cm$^{-3}$). Threading dislocations are dislocations with a component vertical to the lattice-mismatch planes. For the experiments involved, due to the 1D nature of epitaxy growth direction and the profiling technique direction such as SIMS, diffusion in the out-of-plane direction is commonly designed and measured, which is 1D diffusion. For more practical 2D and 3D material systems such as SiGe source and drain regions, diffusion can be calculated using a finite element method and transport equations established and calibrated based-on 1D diffusion data.

## 3.2 Summary of major findings

Table 1 summarizes major discoveries on the Si-Ge interdiffusion impacting factors.

| Topics | Main discoveries |
|---|---|
| Undoped, low-defect and unstrained SiGe interdiffusivity | A unified Si-Ge interdiffusivity model for the full Ge range was built, and verified by data and literature work [94]. <br> (a) $\widetilde{D}$ has a near-exponential relationship with $x_{\text{Ge}}$ at the Si and Ge ends with different slopes. From $x_{\text{Ge}} = 0$ to 1, the interdiffusivity increases by 5 to 6 orders of magnitude in 700 to 900 °C for relaxed SiGe. |

| | (b) 8 to 17X enhancement is observed for every 50 °C increase at the $x_{Ge} = 0$ end, and 4 to 7X enhancement for every 50 °C at the $x_{Ge} = 1$ end in 700 to 900 °C. |
|---|---|
| Compressive strain impact | $\widetilde{D}$ depends exponentially on compressive stain [1, 95]. |
| Tensile strain impact | Not observable for structures with up to 30% Ge with 1% tensile strain [112]. |
| Compressive strain relaxation effect | When the strain relaxation history is known, this effect can be modelled as a smaller strain in the compressive strain term [113]. |
| Threading dislocation impact | It introduces a dislocation-mediated interdiffusion term, which is significant for low Ge regimes [1, 113]. |
| Doping impact | P and As enhances interdiffusion. B has a small impact. N-doping effect can be modelled as a Fermi effect [1, 5, 6]. |
| Defect engineering by carbon | 1.2% carbon enhances interdiffusion of $Si/Si_{0.79}Ge_{0.21}$ superlattices [114]. |
| Defect engineering by oxidation | Not observable for structures with 15 to 60% Ge and a -1% compressive strain [115]. 123% enhancement for $Si_{0.89}Ge_{0.11}/Si_{0.94}Ge_{0.06}$ superlattice [116]. |

Table 1. Summary of the influences from impacting factors of Si-Ge interdiffusion.

The interdiffusivity models and data generated in Xia et al. and Dong et al.'s work [1, 94, 95] have been widely used and implemented in the state-of-the-art process simulation tools including Intel's in-house process simulation tool, Crosslight Software's CSUPREM[TM] and Synopsys's Sentaurus Process[TM] (the leading commercial 3D process simulation tool in the semiconductor industry), and Lumerical's DEVICE[TM] for structure and process design of next generations of semiconductor device.

### 3.3 Biaxial strain impacts

For SiGe systems, Si is commonly used as the substrate. Therefore, SiGe layers are commonly under compressive strains. For MOSFET applications, tensile strains are desired for n-type MOSFETs for electron mobility enhancement and compressive strain are desired for hole mobility enhancement. External stress liners and/or source and drain stressors can be used to introduce strains in the channels.

Xia et al.'s work [112] in 2006 studied the stress impact on the interdiffusion, and observed that tensile stress has little impact, while compressive stress enhances interdiffusion significantly. There have been little studies on the tensile strain impact on interdiffusion other than Ref. [112], which shows that up to 1% biaxial tensile strain has little impact on the interdiffusion where compressive biaxial strains increase interdiffusion significantly (Fig. 17).

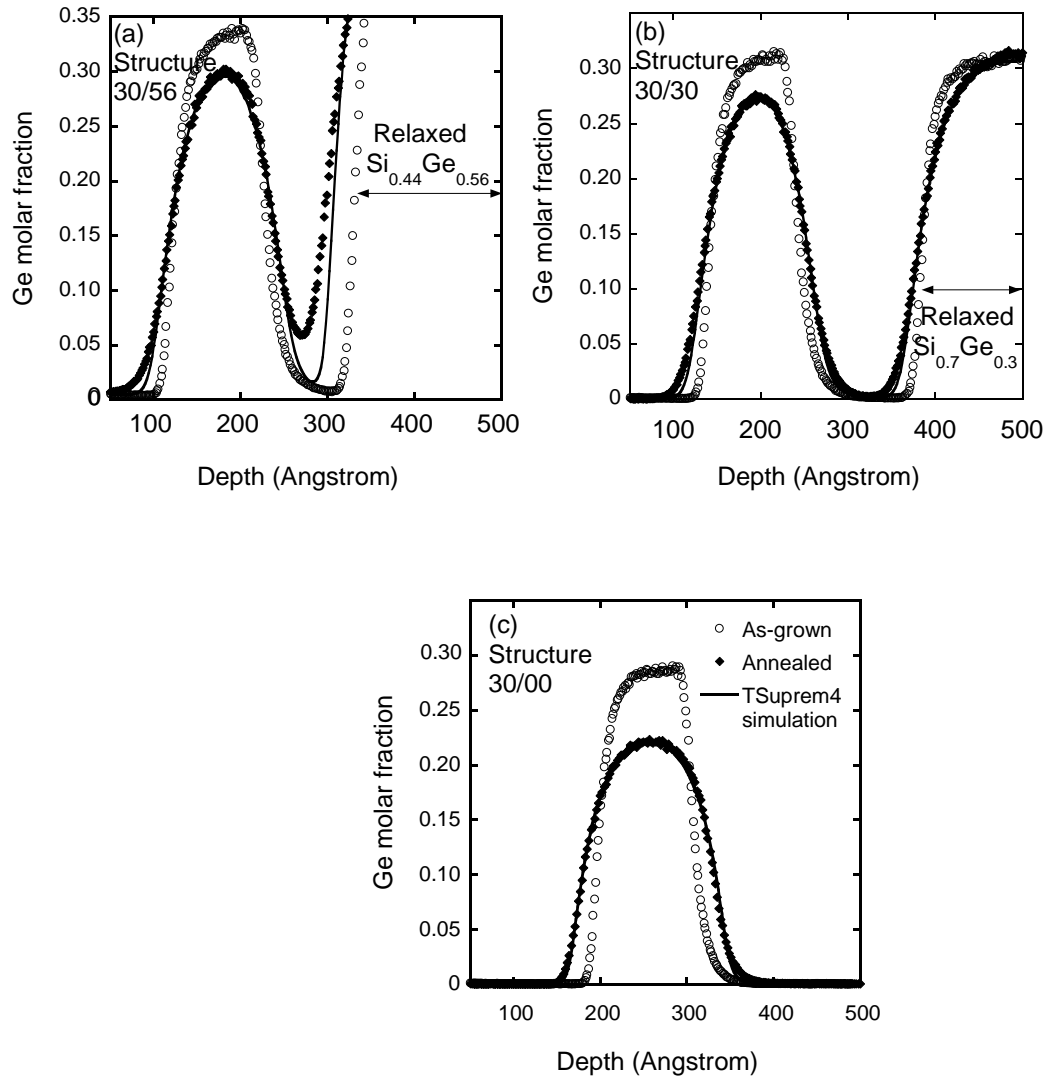

**Fig. 17. As-grown and annealed SIMS profiles of $Si_{0.70}Ge_{0.30}$ peaks on (a) relaxed $Si_{0.44}Ge_{0.56}$, (b) relaxed $Si_{0.70}Ge_{0.30}$, and (c) Si substrates. The annealing condition is 880 °C for 90 min. Comparison of the profiles in (a) and (b) indicates that tensile strain in the $Si_{0.70}Ge_{0.30}$ has little impact on interdiffusion, while compressive strain is associated with a large interdiffusion enhancement (comparison of profiles in (b) and (c)). Figure reproduced from ref. [112] with permission from AIP publishing.**

The stress/strain impact on Si-Ge interdiffusion is an interesting topic. Is the stress gradient an extra driving force for interdiffusion on top of the chemical gradient driving force? Does the stress impact the activation energy and the pre-factor of interdiffusivity? Is the impact from stress gradient or from the stress-related defects? Quantitatively, what is the magnitude of the impact? How can it be modeled? These topics were systematically investigated in ref. [95]. The $x_{Ge}$ studied ranged from 0.36 to 0.75, and the temperature range was 720 to 880 °C. The form of strain is biaxial strain, which is naturally formed in epitaxial SiGe structures. In the industry practice, although stress/strain can be 3D distributions. In a lot of cases, the stress/strain distributions can be approximated as biaxial strain distributions. The epitaxial SiGe structures were kept pseudomorphic during the annealing.

### 3.3.1 Biaxial strain impact on the driving force of interdiffusion

Theoretical analysis in ref. [117] showed that a biaxial strain field adds a term to the interdiffusion driving force on top of the driving force from the chemical concentration gradient. With some derivations using the Gibbs free energy contributions from chemical mixing, elastic biaxial strain, chemical gradient energy and elastic gradient energy, this effect can be included in an apparent interdiffusivity $\widetilde{D}_{apparent}$ as defined below:

$$\widetilde{D}_{apparent} = \widetilde{D}^{strained}\left(1 + \frac{2V_m \eta^2 Y_{UVW}}{G_C''}\right) \quad (19)$$

In the above equation, $\widetilde{D}^{strained}$ is the interdiffusivity under a biaxial strain. $V_m$ is the molar volume of the solid solution, and for SiGe, $V_m \approx 13 \text{ cm}^3/\text{mol}$. $\eta$ is the lattice mismatch. It is 0.0418 for Si and Ge. $Y_{UVW}$ is the biaxial modulus, which equals $E/(1-\upsilon)$. E is Young's modulus, and $\upsilon$ is the Poisson ratio. $G_C''$ is the second derivative of the Gibbs free energy term for chemical mixing over $x_{Ge}$. The term $\frac{2V_m \eta^2 Y_{UVW}}{G_C''}$ reflects the magnitude of the biaxial strain energy contribution to the driving force, which depends on $x_{Ge}$ and temper100ature. Fig. 18 has the strain field effect term $\frac{2V_m \eta^2 Y_{UVW}}{G_C''}$ as a function of $x_{Ge}$ and temperature. For $x_{Ge} = 0.5$ at 1000 K, for example, this term can be as large as 0.33, so the strain contribution to the driving force should not be neglected [114, 115]. When $x_{Ge}$ is close to 0 or 1 at high temperatures, this factor is negligible.

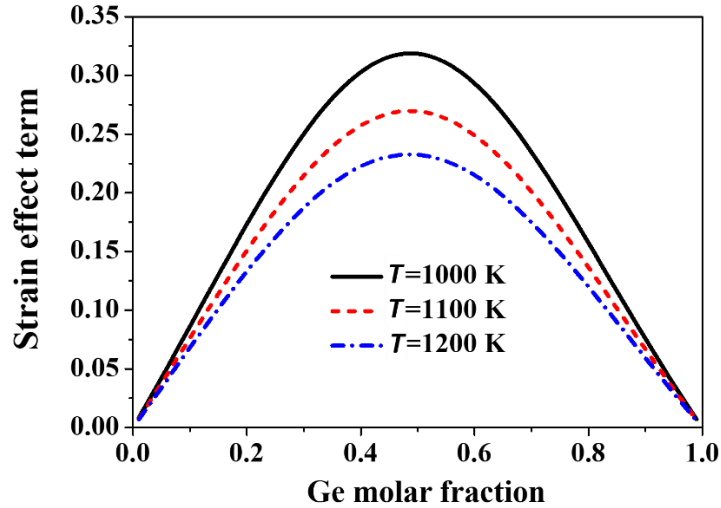

**Fig. 18. Ge concentration dependence of the strain effect term $\frac{2\eta^2 Y_{UVW}}{G_C''}$ at different temperatures.**

For epitaxial SiGe systems with $<UVW> = <100>$, it can be shown that:

$$\widetilde{D}_{apparent} = \left(1 + \frac{2 \times 0.0418^2 Y_{100}}{f_0''}\right)\widetilde{D}^{strained}. \quad (20)$$

In the above equation, $Y_{100}$ is the biaxial modulus of (100) orientation Si$_{1-x}$Ge$_x$; $f_0(x_{Ge})$ is the Helmholtz free energy per unit volume of the homogeneous solution, and $\widetilde{D}^{relaxed}$ is the interdiffusivity with no strain. The calculation of $f_0''$ can be found in [95].

### 3.3.2 Biaxial strain impact on the interdiffusivity

Besides the change in the driving force, strain can change $\widetilde{D}^{strained}$ itself. The definition of the strain derivative of the interdiffusivity, $q'$, is shown in Equation (21). It includes the effects of strain on the diffusivity prefactors and on the activation energy.

$$q'(T) \equiv -\left(\frac{kT}{\varepsilon}\right) \ln\left(\frac{\widetilde{D}^{strained}}{\widetilde{D}^{relaxed}}\right) \quad (21), \text{ and thus}$$

$$\widetilde{D}^{strained} = \widetilde{D}^{relaxed} e^{\frac{-q'\varepsilon}{kT}} \quad (22)$$

Taken the strain relaxation factor $R$ into account, the apparent interdiffusivity can be expressed as

$$\widetilde{D}_{apparent} = \left(1 + \frac{2\eta^2(1-R)^2 Y_{UVW}}{f_0''}\right) \widetilde{D}^{relaxed} e^{\frac{-q'\varepsilon_0(1-R)}{kT}}. \quad (23)$$

$\varepsilon_0$ is the strain without any relaxation. $q'$ was quantitatively extracted from the experimental data in [95] as

$$q' = (-0.081T + 110) \text{ eV}. \quad (24)$$

Therefore, the interdiffusivity under a biaxial compressive strain can be expressed as

$$\widetilde{D}^{strained} = \widetilde{D}^{relaxed} e^{\frac{-q'\varepsilon}{kT}} = \widetilde{D}^{relaxed} e^{\frac{(0.081T-110)eV*\varepsilon}{kT}}. \quad (25)$$

Table 2 summarizes all the $q'$ value from literature. Dong et al.'s work [95] extended the Ge fraction range to 0.75, and $q'$ value from that work is comparable to other $q'$ values reported at lower Ge fraction ranges. There are little systematic data on the $x_{Ge}$ dependence of $q'$. Overall, $q'$ is in the range of 10 to 45 eV/unit strain. More experimental studies for different Ge fractions and temperatures are needed.

| References | $x_{Ge}$ | Maximum compressive strain $\varepsilon$ | Temperature (°C) | $q'$ (eV/unit strain) |
|---|---|---|---|---|
| Cowern et al. [87] | 0.25 | -1.05% | 900 to 1050 | 35 to 45 |
| Cowern et al. [118] | 0.3 | -1.26% | 875 | 13.5 to 22.1 |
| Aubertine et al. [119] | 0.17 | -0.71% | 795 to 895 | 19 |

| Aubertine et al. [96] | 0.075 to 0.19 | -0.81% | 770 to 870 | 10 |
| Xia et al. [1] | 0.30 to 0.56 | -1.05% | 770 to 920 | 27.6 to 36.3 |
| Dong et al. [95] | 0.36 to 0.75 | -1.20% | 720 to 880 | 16.6 to 29.6 |

**Table 2. Experimental values of the strain derivative of the interdiffusivity $q'$ in Si-Ge interdiffusion.**

$q'$ includes the impact of strain on both Ge and Si diffusion. For Dong et al's samples [95] with medium Ge fraction, not only $x_{Ge}$ and 1- $x_{Ge}$, but also Ge and Si self-diffusivities are comparable, so the impact of strain on Ge diffusion and Si diffusion both contribute to $q'$. As is known, on the atomic scale, Si and Ge diffusion in SiGe can be mediated by either interstitials or vacancies, or both. Theiss et al. [120] and Aziz [121] has shown that compressive strain enhances vacancy mediated diffusion while retards interstitial mediated diffusion. Therefore, $q'$ for Si-Ge interdiffusion should also depend on the I- and V-mediated fractions, which depend on both temperature and Ge fraction. In Si and Si rich SiGe, Si and Ge self-diffusion are mediated by both interstitials and vacancies [66, 103, 106, 122, 123].

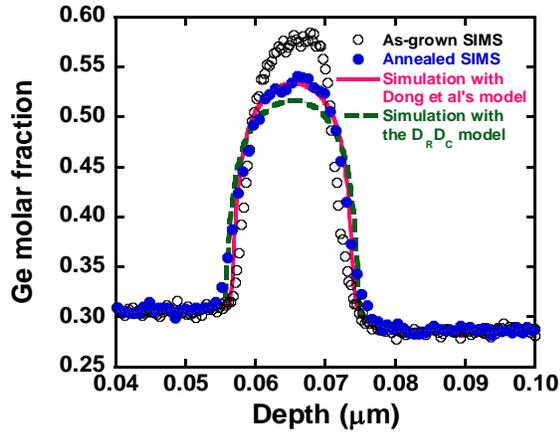

**Fig. 19 As-grown, annealed Ge SIMS profiles of a $Si_{0.70}Ge_{0.30}/Si_{0.45}Ge_{0.55}/Si_{0.70}Ge_{0.30}$ heterostructure and predictions simulated by Dong et al.'s model [94] and the $D_R D_C$ model [2]. The furnace annealing was performed at 800 °C for 40 min. Figure reproduced from ref. [123] with permission from Taylor and Francis Group.**

Compared with the $D_R D_C$ model in Eqs. (12) and (13), which is a good empirical model, the compressive strain impact in Dong et al.'s 2014 work [95] is more scientifically sound in terms of thermodynamics and diffusion theory, and should work for a larger temperature range. In terms of practical applications, both models give reasonable predictions within the error bar. Predictions from both models are compared with SIMS data, shown in Fig. 19.

### 3.4 Doping impacts
For HBT applications, the Ge molar fraction is commonly below 0.3, and SiGe layers are typically tens of nanometer thick and under compressive strain. P and B were shown to enhance Si-Ge interdiffusion in Si/Si$_{0.8}$Ge$_{0.2}$ (Fig. 20) [114]. Carbon is not a dopant, but it also increases the interdiffusion as seen in Fig. 20d.

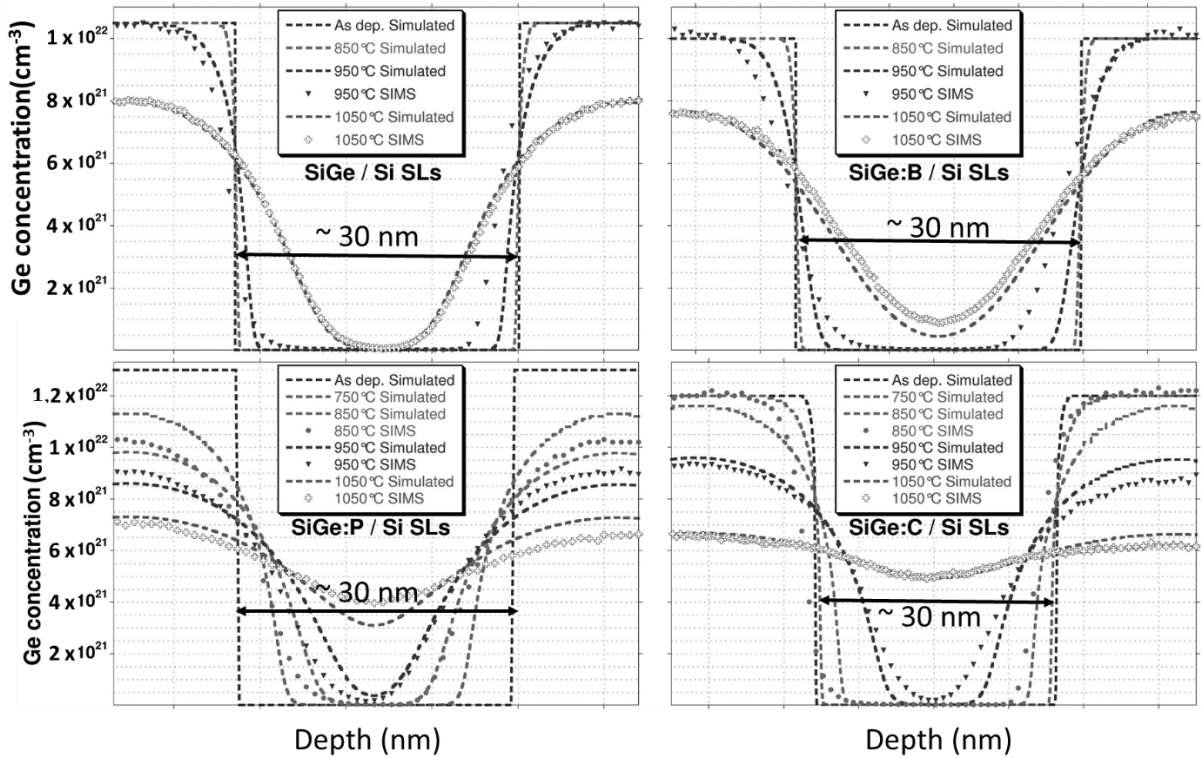

**Fig. 20. TOF-SIMS as-deposited and annealed Ge profiles of the first valley of SiGe/Si, SiGe:B/Si, SiGe:P/Si and SiGe:C/Si superlattices (SL). Figure reprinted from ref. [114] with permission from AIP publishing.**

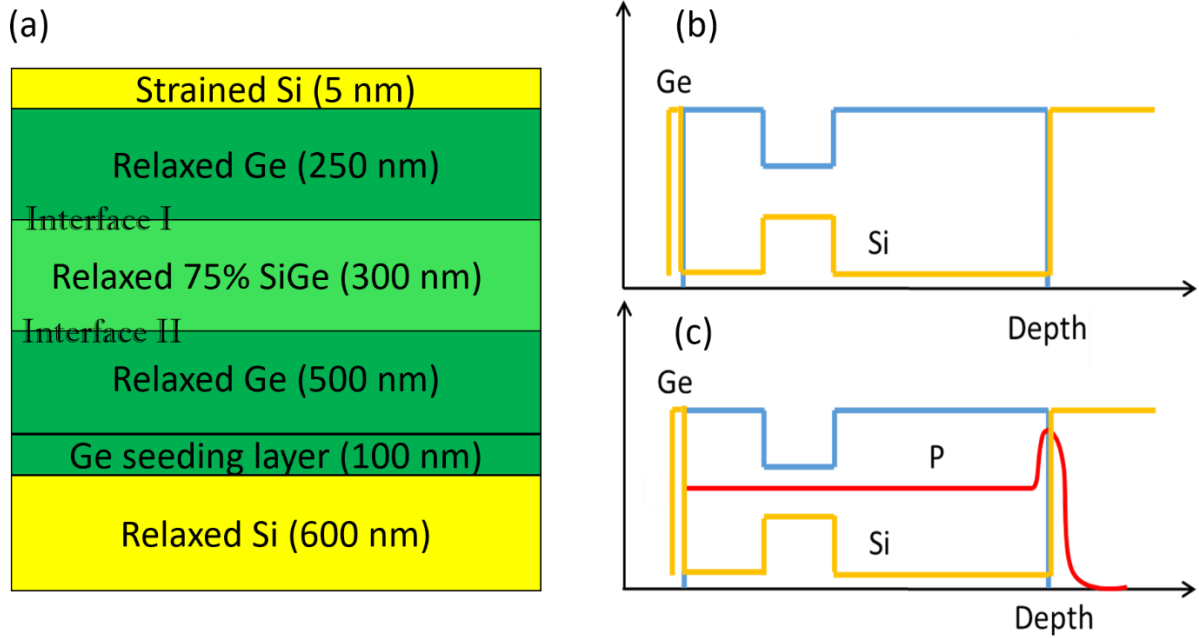

**Fig. 21. Samples used the doping effect study: Sample structure and growth temperature (a), depth profile of the sample with no P doping (b) and depth profile of the sample (c) with P doping concentration at around $5\times10^{18}$ cm$^{-3}$. Figure reprinted from ref. [5] with permission from AIP publishing.**

For Ge-on-Si laser applications, Ge is close to 100% Ge. The thickness is commonly a few hundred of nanometers resulting in fully relaxed Ge films on Si. A high n-type doping is typical and the interdiffusion has a significant impact on the light emission processes. In terms of the chemical composition, Ge-on-Si forms an interdiffusion couple that is a convenient structure for interdiffusion study. However, common epitaxial Ge/Si interfaces are highly defected due to the Ge seeding layers that are grown to reduce the dislocation density of the top Ge layers. Dopant diffusion, dopant segregation and interdiffusion are coupled and hard to separate or measure by SIMS. Therefore, Ge-on-Si structures are not ideal for interdiffusion studies.

To address this problem, Ref. [5] investigated the P doping effect on interdiffusion with Ge/Si$_{1-x}$Ge$_x$/Ge on Si structures (Fig. 21) such that the highly-defected Ge/Si interfaces are sufficiently away from the interdiffusion region of our interest. The samples have $0.75 < x_{Ge} < 1$ and a mid-$10^{18}$ to low-$10^{19}$ cm$^{-3}$ P doping. The dislocation densities were around $10^8$ to $10^9$ cm$^{-2}$ range. The P-doped sample shows an accelerated Si-Ge interdiffusivity, which is 2 to 8 times of that in the undoped sample. As Interface I and II, the interdiffusion regions of our interests, are not at the Ge/Si interfaces, the dislocation density in this study was assume to be not significant, and the dislocation mediated diffusion was ignored for this case. The doping dependence of the Si-Ge interdiffusion was modelled by a Fermi-enhancement factor "FF".

$$\widetilde{D}_{\text{total}} \approx \widetilde{D}_{\text{lattice,undoped}} * FF = \widetilde{D}(n=n_i) * FF \qquad (26)$$

$$FF \equiv \frac{\widetilde{D}(n)}{\widetilde{D}(n_i)} = \frac{1+m_1 \exp\left(\frac{E_i-E_{V^-}}{kT}\right)\left(\frac{n}{n_i}\right)+m_2\exp\left(\frac{2E_i-E_{V^-}-E_{V^{2-}}}{kT}\right)\left(\frac{n}{n_i}\right)^2}{1+m_1 \exp\left(\frac{E_i-E_{V^-}}{kT}\right)+m_2\exp\left(\frac{2E_i-E_{V^-}-E_{V^{2-}}}{kT}\right)} \qquad (27)$$

Here, we denote

$$\beta = m_1 \exp\left(\frac{E_i-E_{V^-}}{kT}\right), \qquad (28)$$

$$\gamma = m_2 \exp\left(\frac{2E_i-E_{V^-}-E_{V^{2-}}}{kT}\right), \qquad (29)$$

where $E_i, E_{V^-}$ and $E_{V^{2-}}$ are the electron energy levels for intrinsic SiGe, single negatively charge vacancies and double negatively charged vacancies. The experiments show that Si-Ge interdiffusion coefficient is proportional to $n^2/n_i^2$ for $0.75 < x_{Ge} < 1$. This indicates that the interdiffusion in high Ge fraction range with n-type doping is dominated by $V^{2-}$ defects. $FF$ was shown to have a relatively weak dependence on the temperature and $x_{Ge}$. $m_1 = 1$ and $m_2 \geq 20$ generate good fitting to the experimental data (Fig. 22).

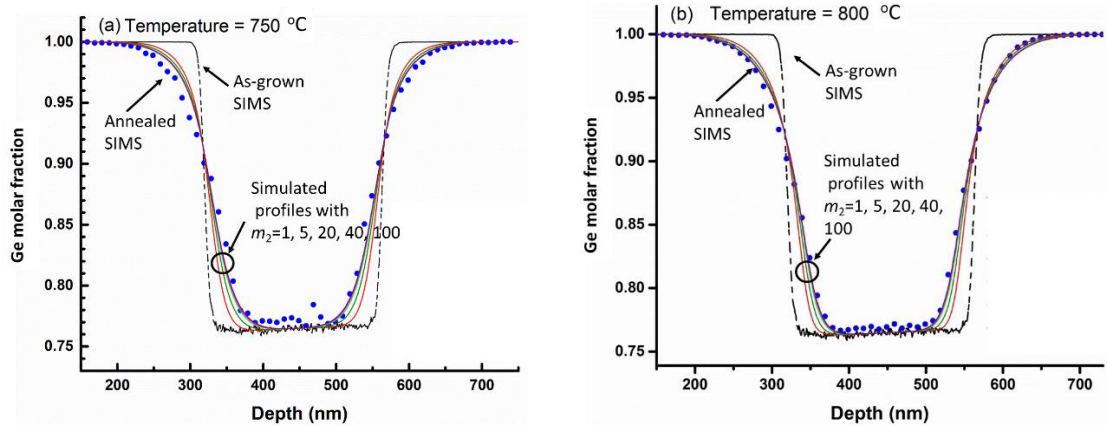

**Fig. 22** Comparison between SIMS data and calculations using Equation (26) to (29). Ge profile fitting (a) at 750 °C for 120 min; (b) at 800 °C for 30 min. $m_1$ is fixed to 1 in each simulation and $m_2$ is 1, 5, 20, 40, 100 separately. The diffused profiles have a weak dependence on $m_2$. Figure reprinted from [5] with permission from AIP publishing.

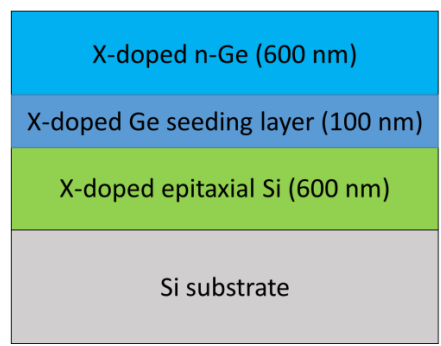

**Fig. 23** Epitaxial structure used in [6] to study the doping impact on material quality and Si-Ge interdiffusion. X stands for As, higher boron, lower boron, phosphorus-doped or undoped. Figure modified from [6], an open access paper by OSA Publishing.

Ref. [6] further investigated Ge-on-Si epitaxial film quality and interdiffusion with three different dopants (P, As and B) and those without intentional doping for the full Ge range (Fig. 23). Some have seen a high temperature (nominal 850 °C) and low temperature (nominal 680 °C) thermal cycling defect annealing step, while some have not. The doping levels after defect annealing can be found in Fig. 24(b). All samples have a smooth surface (roughness < 1.5 nm), and the Ge films are almost entirely relaxed. Etch pit density (EPD) data have been obtained for the doped Ge, which shows that boron-doped Ge has the highest EPD in $10^8$ cm$^{-2}$.

Caution needs to be taken when using EPD as a TDD characterization method [124]. Although EPD methods have been used to give rough estimations of TDD in undoped Ge, it can generate large discrepancies for doped Ge due to the change in the etching property, where a significant portion of TDD do not show as etch pits. In these cases, electron channeling contrast imaging (ECCI) can be used, which is a SEM-based non-destructive characterization technique [124]. In ECCI, the intensity of electrons backscattered from the surface of the sample under investigation is imaged [124]. Slight deformations of the crystal lattice, such as the strain field associated with dislocations, lead to a strong modulation of the backscatter intensity which can be observed using a backscatter detector. Differences between EPD and ECCI results can be orders of magnitude. For example, for P-doped Ge without defect annealing, EPD measured was $3 \times 10^5 \ cm^{-2}$, while ECCI measured TDD was $5.2 \times 10^8 \ cm^{-2}$. These TDD are measured from the top surface of Ge, which is a good indication of the Ge film quality.

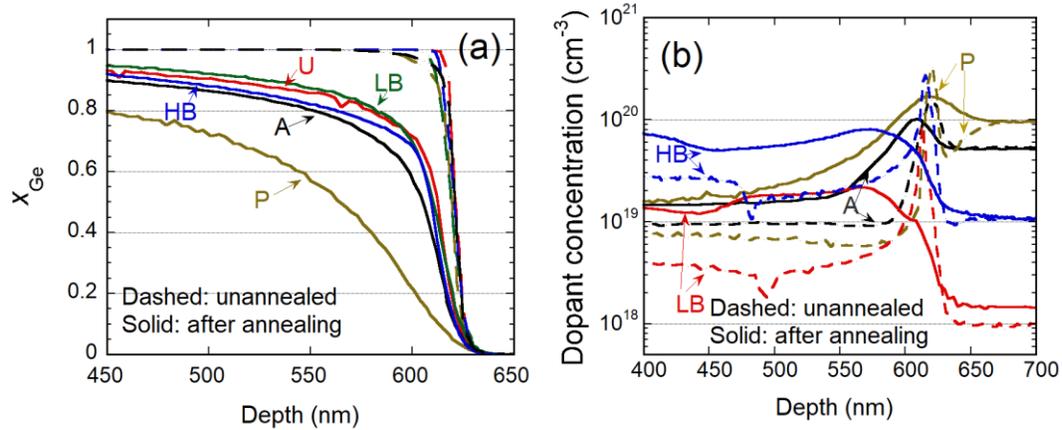

**Fig. 24. Doping impacts on (a) Ge profiles and (b) dopant profiles measured by SIMS. A, HB, LB, P and U stand for As, higher boron, lower boron, phosphorus-doped or undoped Ge/Si sample. Figure reprinted from [6], an open access paper by OSA Publishing.**

Before annealing, all samples have very similar sharp Ge profiles (Fig. 24) and dopants have the highest concentration peaks at the interface of Ge/Si. This is due to the segregation induced by a high density of defects in the Ge seeding layers. Extracted effective interdiffusivity with different doping are shown in Fig. 25. Sample P has the most interdiffusion. Sample A has the second largest interdiffusion. Sample U and sample LB have the least interdiffusion. For sample HB, it has no significant difference over sample LB in $x_{Ge} < 0.7$ part, but it distinguishes itself from LB and U in $x_{Ge} > 0.7$ part. The interdiffusion profiles are more box-like than Gaussian profiles, showing a strong dependence.

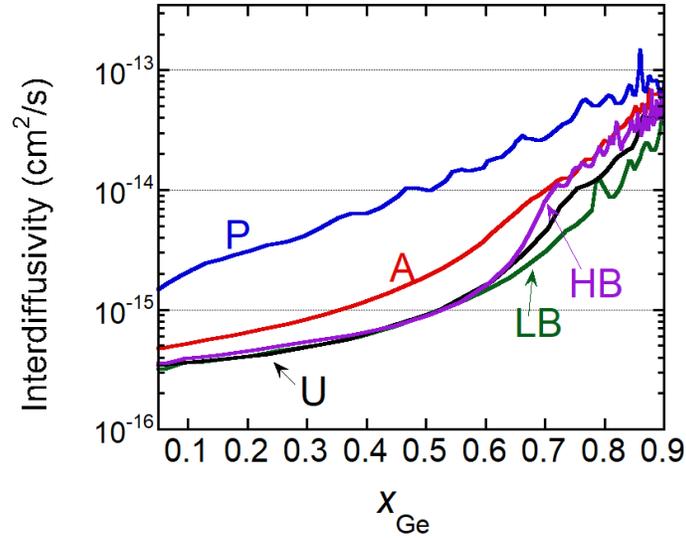

**Fig. 25.** Extracted effective interdiffusivity with different doping. A, HB, LB, P and U stand for As, higher boron, lower boron, P-doped and undoped Ge/Si sample respectively. Figure reprinted from [6], an open access paper by OSA Publishing.

In the interdiffusion region of the interest, which is at Ge seeding layer/Si interfaces, dislocation density is much higher due to the high lattice mismatch. Therefore, unlike the study in [5], the dislocation mediated term cannot be ignored in the modeling. Therefore, the total interdiffusivity is expressed as the following.

$$\widetilde{D}_{\text{total}} = \widetilde{D}_{\text{dislocation}} + \widetilde{D}_{\text{lattice,undoped}} * FF. \tag{30}$$

Due to limited data of $n_i(x_{Ge})$, Cai et al. [5] used a linear interpolation between $n_{i,Ge}$ and $n_{i,Si}$ as in the equation below, which is good enough for $0.75 < x_{Ge} < 1$.

$$n_i(x_{Ge}) = n_{i,Ge} x_{Ge} + n_{i,Si}(1 - x_{Ge}) \tag{31}$$

However, Zhou et al.'s work covers $0 < x_{Ge} < 1$ range. According to ref. [125], Zhou et al. [6] compared the linear and the exponential interpolation between $n_{i,Ge}$ and $n_{i,Si}$ as shown in Equation (32):

$$n_i(x_{Ge}) = n_{i,Si} \exp(ln \frac{n_{i,Ge}}{n_{i,Si}} \times x_{Ge}). \tag{32}$$

It was shown that the exponential relation between the intrinsic carrier concentration $n_i$ and $x_{Ge}$ gives better Ge profile fitting results for P-doped and As-doped Ge/Si [6].

On the light emitting properties of doped Ge, photoluminescence (PL) measurements show that P and As-doped Ge without defect annealing show a 5 to 10 times enhancement in PL intensity owing to the fact that the interdiffusion is minimized for unannealed samples, which have higher TDD but less Si-Ge intermixing [7].

### 3.5 Defect impacts

In crystalline materials with low-density 2-dimensional or 1-dimensinal defects such as stacking faults, grain boundaries and dislocations, diffusion and interdiffusion are mediated by point defects.

Point defect engineering has been quite successful in controlling dopant diffusion. One example is the use of carbon to retard boron diffusion in Si and SiGe for the SiGe HBT industry. In order to find ways to control interdiffusion, it is very important to understand the interdiffusion mechanisms.

Thermal oxidation of silicon and carbon incorporation are both considered as major approaches in engineering point defect concentrations [131]. The former one injects interstitials and reduces vacancy concentrations and the latter one consumes interstitials and increases vacancy concentrations [126]. The oxidation impact on interdiffusion was investigated by Ozguven and McIntyre [116] and Xia and Hoyt [115], which rendered different results. The former one studied $Si_{0.89}Ge_{0.11}/Si_{0.94}Ge_{0.06}$ superlattices coherent to a Si substrate, which showed a 123% enhancement in the interdiffusivity at 795 °C during oxidation. This supports that interdiffusion has an interstitial-mediated component. The latter one studied compressively strained $Si_{0.41}Ge_{0.59}$ and $Si_{0.56}Ge_{0.44}$ layers with about -1% compressive strain. Interdiffusion under an effective inert, i.e., underneath masking layers, and oxidizing conditions showed no differences. This discrepancy may come from the fact that these two studies dealt with different Ge fractions, i.e., an average 8.5% Ge compared to 44% to 59% Ge. It is likely that the interstitial concentration injected by oxidation had an effect on the lower Ge concentration, but was too low to impact SiGe alloys with much higher Ge concentrations.

The effect of carbon is shown Fig. 20 (d) above. The carbon used has a total concentration of $6 \times 10^{20} cm^{-3}$, and the substitutional carbon was about $4.6 \times 10^{20} cm^{-3}$. The total carbon level is 1.2 at.%, which is much larger than the common carbon level of less than 0.3 at.% used in SiGe HBTs. Nevertheless, this carbon impact supports a vacancy-mediated interdiffusion mechanism. Combined with the experiment results in ref. [116], it is reasonable to believe that Si-Ge interdiffusion is mediated by both interstitials and vacancies.

Besides the point defects, dislocations are common 1D defects in epitaxial SiGe systems. They serve as fast interdiffusion paths, which add a dislocation-mediated interdiffusion term as discussed previously and are not desired.

A technically very significant question to address is on how to retard the interdiffusion, which is one of the most important purposes of the interdiffusion mechanism and defect engineering studies. As discussed above, thermal oxidation and carbon incorporation were investigated, but they were not effective in retarding interdiffusion. The reasons may be two-fold. First, for Si-Ge interdiffusion, the concentration of Si and Ge (on the order of $10^{22}$ cm$^{-3}$) are much higher than dopant concentrations (between $1 \times 10^{15}$ to $2 \times 10^{20}$ cm$^{-3}$). Second, interdiffusion is believed to be mediated by interstitials and vacancies with vacancy-mediated interdiffusion as the dominant term. This dual diffusion mechanism is not desired for interdiffusion control as point defect engineering is most effective when one type of defects is responsible for (inter)diffusion. To the author's best knowledge, there have not been any reports on effective ways in controlling interdiffusion in group IV alloys and heterostructures.

# 4 Summary and prospective

As discussed above, group IV alloys and heterostructures have been used extensively in the semiconductor industry for over three decades. Interdiffusion changes the chemical concentration distributions and thus many material properties and device performance. Over the past years, significant efforts have been devoted to answer this seemingly basic and fundamental question: for Si and Ge, the two very important and closely related semiconductors, how do they intermix. Getting the answers has taken many years when we tried to answer it in the atomic scale, as discussed above.

Although controlling the interdiffusion is an even harder topic, the interdiffusion studies so far have been very useful to deal with the interdiffusion problem in the design stage. Many interdiffusion cases have been modelled reasonably well, which can be predicted using simulation tools such as Sentaurus Process$^{TM}$ by Synopsys, CSUPREM$^{TM}$ by Crosslight Software and DEVICE$^{TM}$ by Lumerical Solutions. The structures, compositions and processing conditions can then be designed such that the amount of interdiffusion is taken into account during device fabrication processes.

So far, Si-Ge interdiffusion has been studied from a general point of view, where large ranges of impacting factors were studied. More detailed studies can be done for specific applications in the future. The major constraint of interdiffusion studies lies in the constraint of the 1D profiling technique and its accuracy. Many industry structures have 3D strain fields and 3D element distributions. Although 3D predications can be done with the interdiffusion models and parameters calibrated with 1D structures. SIMS as the most widely used 1D profiling metrology for interdiffusion studies has its own limitations such as the SIMS broadening effect at interfaces. Also, the depth resolution of SIMS is commonly around tens of nm per decade, which is not very accurate for atomic scale profiling. Atom probe tomography (APT) can obtain 3D element distributions, but so far it has been prohibitively expensive. Interdiffusion behaviors in Ge-Sn systems are much less studied, which are also topics for future work.

# 5 Conflict of interest

The author declares that she has no conflict of interest.

# 6 Acknowledgments

The author would like to acknowledge Natural Science and Engineering Research Council of Canada (NSERC), Crosslight Software Inc. and Lumerical Solutions for funding the Si-Ge interdiffusion studies at UBC. Crosslight Software Inc. and Lumerical Solutions are acknowledged for their technical assistance in implementing the interdiffusion models in their simulation tools CSUPREM$^{TM}$ and DEVICE$^{TM}$. The epitaxial structures of these studies were grown by Gary Riggott at the Microsystems Technology Laboratories at MIT, and Dr. Kwang Hong Lee in Prof. Chuan Seng Tan's group at Nanyang Technological University in Singapore. The Nanofabrication facility at UBC, Mattson Technology Canada and the Semiconductor Defect

Spectroscopy Laboratory at Simon Fraser University are acknowledged for the technical assistance in the material processing and characterizations. CMC Microsystems is acknowledged for their support in Synopsys's Sentaurus Process$^{TM}$, TSUPREM-4$^{TM}$ and CUPREM$^{TM}$ licenses and technical support. Dr. Stephen P. Smith at Evans Analytical Group is acknowledged for helpful discussions on SIMS analysis. Tao Fang from the Department of Materials Engineering at UBC is acknowledged for his help in proof-reading this paper.

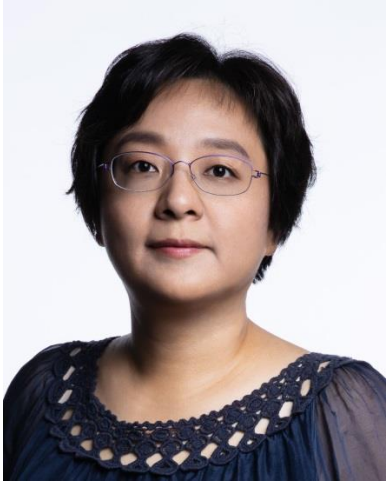

Guangrui (Maggie) Xia received her M.S. and Ph.D. degrees in Electrical Engineering from the Massachusetts Institute of Technology. She is an associate professor in the Department of Materials Engineering at the University of British Columbia, Vancouver, Canada. She has extensive research experiences in SiGe materials and devices. In recent years, her research interests have expanded to Si-compatible lasers, wide bandgap semiconductors, 2D materials, and through-Si-vias for 3D integration of ICs.